\documentstyle[prb,aps]{revtex}

\begin{document}

\draft


\title{Quasiparticle Bound States and Low-Temperature Peaks of
the Conductance of NIS Junctions in d-Wave Superconductors}


\author{Yu.\ S.\ Barash and A.\ A.\ Svidzinsky}
\address{
 P.\ N.\ Lebedev Physics Institute, Leninsky Prospect 53,
 Moscow 117924, Russia}
\author{H.\ Burkhardt}
\address{
 Physikalisches Institut, Universit\"at Bayreuth,
 D-95440 Bayreuth, Germany\\ }


\date{\today}


\maketitle


\begin{abstract}
Quasiparticle states bound to the boundary of anisotropically paired
superconductors, their contributions to the density of states and
to the conductance of NIS junctions are studied both analytically
and numerically. For smooth
surfaces and real order parameter
we find some
general results for the bound state energies. In particular,
we show that under fairly general conditions quasiparticle states
with nonzero energies exist for momentum directions within a
narrow region around the surface normal. The energy dispersion
of the bound states always
has an extremum for the direction along the normal.
Along with the zero-bias anomaly
due to midgap states,
we find, for quasi two-dimensional materials,
additional low-temperature peaks
in the conductance of NIS junctions
for voltages determined by the extrema of the bound state energies.
The influence of interface roughness
on the conductance
is investigated within the framework of Ovchinnikov's model.
We show that nonzero-bias peaks at low temperatures may give
information on the order parameter in the bulk, even though it
is suppressed at the surface.
\end{abstract}





\section{Introduction}

It is now widely recognized, that the anisotropy ratio of the
order parameter on the Fermi surface for many of the high-$T_c$
superconductors is of the order of unity. By contrast, for conventional
low-temperature superconductors
the anisotropy ratio is known to be much smaller, so that they
are almost isotropic $s$-wave superconductors.
This fact itself, essentially irrespective of the specific type of the
superconducting pairing, results in many new important consequences
for the properties of HTSC, compared to isotropic $s$-wave
superconductors. In particular, a highly anisotropic order parameter
turns out to be quite sensitive to any kind of inhomogeneities in
the material, including nonmagnetic impurities as well as interfaces.

For most
surfaces
the bulk behavior of an
anisotropic order parameter can differ essentially from its behavior at
the wall. Even smooth specularly reflecting walls are, in general,
pair breaking, which results in a suppression of the order parameter
near the boundary.
This concerns both anisotropic superconductors
and superfluids.\cite{Amb74,buch81,kur,bgz,mat,nag,buch95}
At the same time, quasiparticle states bound to the boundary appear
due to this suppression or/and due to the
sign change of the anisotropic order parameter with momentum
direction.\cite{buch81,mat,nag,buch95,hu,yhu,bs1} Furthermore,
in the region where the bulk order parameter is essentially suppressed,
subdominant pairing channels
with different symmetries may come into play
and become stable near the wall.\cite{buch95,mat}

Several important experimental methods used for studying the anisotropic
structure of the order parameter, in particular tunneling measurements,
are fairly sensitive to the superconducting properties close to the
surface of the sample.
The local quasiparticle spectrum at the surface which differs,
in general, from the bulk density of states is of crucial importance
for the I-V characteristics of tunnel junctions.
Apart from the fact, that the gap anisotropy washes
out those peaks in the tunneling current which would occur
in the isotropic case, the presence of
quasiparticle bound states or/and the order parameter of subdominant
symmetry channels near the surface results in new characteristic
features of the I-V curves in the vicinity of
some specific points.\cite{tan1,buch95,bs1}
One of the most striking features of NIS junctions with $d$-wave
superconductors is the zero-bias anomaly of the conductance
which is associated with surface midgap states.\cite{tan1}
These states localized at the surface are a consequence of the
sign change of the order parameter on the Fermi surface.
A further consequence of those states is
a low-temperature anomaly in the
Josephson critical current.\cite{tan2,bbr,burk97}
By contrast, some other important features
of the Josephson effect do not depend on the
quasiparticle spectrum at the surface, having a more universal
character.
Actually, they depend only on the phase difference of
the order parameters in the bulk superconductors
on both sides of the junction.\cite{lar,sig,har}

Below we develop an analytical investigation, including numerical
calculations, of quasiparticle states bound to
the boundary of anisotropically paired superconductors,
their contributions to the density of states and to the conductance of
NIS junctions with weak transparency. Our approach to the problems is
based on the quasiclassical formulation of
superconductivity.\cite{eil68,lar68,lar75,eli71,lar86,rai83}
Both smooth specularly
reflecting and diffusive interfaces are considered. The spatially
dependent order parameter is supposed to be real.
The analytical study of quasiparticle states with nonzero
energies bound to a smooth wall is carried out for the first time,
and some general results are obtained.
In particular, we show that the energy of a bound state is closely
connected with the phase of the complex residue
of the anomalous propagator. Quasiparticle bound states with nonzero
energies turn out to exist under fairly general conditions for momentum
directions within a narrow region around the surface normal
for which the energy dispersion always has an extremum.

While the presence of a zero-bias anomaly in the conductance does not
depend upon
details of the Fermi surface
but only on the
appearance of midgap states,
low-temperature anomalies in the conductance due to
nonzero-energy levels near the surface are absent for three dimensional
systems. For quasi two-dimensional materials we find, that quasiparticle
bound states with nonzero energies
lead, at least at low enough temperatures, to peaks in the
conductance for voltages determined by the extrema
of the bound state energies.
It is shown below, that for a smooth barrier plane 
the magnitudes of those peaks
are proportional
to the inverse
square root of the temperature
($G(T \rightarrow 0) \propto 1/\sqrt{T}$)
in contrast to
zero-bias anomalies for which
$G(T \rightarrow 0) \propto 1/T$.

In realistic systems, the peaks in the quasiparticle spectrum
will be broadened. The shape and the height of these peaks are governed
by the physical parameters of the junction
like interface roughness (e.g., lattice constant mismatch)
and transparency.
We investigate the influence of interface roughness on the
local density of states\cite{mat,nag96b} and on the conductance
within the framework of Ovchinnikov's model.\cite{ovc69,cul84,burk97}
In all our calculations the order parameter is determined
self-consistently.
In contrast to midgap states, nonself-consistent
calculations, disregarding surface pair breaking, do not permit to find
and describe quasiparticle bound states with nonzero
energy and the corresponding peaks in the conductance at nonzero-bias
voltage.

\section{Quasiclassical Theory}

\subsection{Basic Equations}
In general, strongly anisotropic superconductors react sensitively
to surfaces and interfaces.
For instance, the superconducting order parameter
near the surface is reduced due to pair-breaking effects.
The quasiclassical theory of superconductivity
is the most efficient theory for investigating effects of surfaces and
interfaces.
We use Eilenberger's equations
for the quasiclassical retarded propagator $\hat{g}^R$
which for a clean singlet anisotropically
paired superconductor reduce to the following
$2\times 2$ matrix form:\cite{not}
\begin{eqnarray}
\left[\varepsilon\hat{\tau}_3-\hat{\Delta}(\bbox{p}_f,\bbox{R}),
\hat{g}^R(\bbox{p}_f,\bbox{R};\varepsilon)\right] +
i \bbox{\rm v}_f\hspace*{-2pt} \cdot \hspace*{-2pt}
\bbox{\nabla} \hspace*{-2pt}_{\bbox{R}}\:
\hat{g}^R(\bbox{p}_f,\bbox{R};\varepsilon)=0 \enspace, \label{eil} \\
\big[\hat{g}^R(\bbox{p}_f,\bbox{R};\varepsilon)\big]^2=
-\pi^2\hat{1} \enspace.
\qquad\qquad\qquad
\label{norm}
\end{eqnarray}
Here, $\varepsilon$ is the quasiparticle energy,
$\bbox{p}_f$ the momentum on the Fermi surface,
$\bbox{\rm v}_f$  the Fermi velocity, and
$\hat\Delta$ the order parameter matrix. Following standard notations we
use a ``hat'' to indicate matrices in Nambu space.
A convenient basis set of
Nambu matrices is the unit matrix $\hat 1$ and the three Pauli matrices
$\hat \tau_1$, $\hat\tau_2$, $\hat\tau_3$. The propagator $\hat g$
(here and in the following we drop the superscript $R$ for simplicity)
and the order parameter matrix $\hat\Delta$ have the form
\begin{eqnarray}
\hat g=\left(\begin{array}{cr}
                                   g   &  f \\
				   f^+ & -g
	    \end{array}\right)\enspace
\qquad \mbox{and} \qquad
\hat \Delta=\left(\begin{array}{cc}
                                       0     &  \Delta \\
				   -\Delta^* &     0
	    \end{array}\right)\enspace.
\label{matrix}
\end{eqnarray}
The gap function $\Delta(\bbox{p}_f,\bbox{R})$ is related to the
anomalous Green function $f$ and has to be determined self-consistently
(see Ref.\ \onlinecite{buch95}). The diagonal part $g$ of the
full matrix propagator $\hat g$ carries information on the
superconducting density of states (DOS),
\begin{eqnarray}
\nu(\bbox{p}_f,\bbox{R};\varepsilon)=
-\frac{1}{\pi}{\rm Im}\big[g(\bbox{p}_f,\bbox{R};\varepsilon)\big]
\enspace.
\label{dos}
\end{eqnarray}
This quantity contains all the
information we need for calculating the tunneling current
across an NIS interface (see Eqs.\ (\ref{j})).

\subsection{Interface Model}
The boundary conditions at an ideal interface are given by Zaitsev's
relations (see Refs.\ \onlinecite{zai,kur}),
which reduce in the limit of zero-transparency to \cite{kul}
\begin{equation}
\hat {g}(\bbox{p}_f)=\hat{g}(\underline{\bbox{p}}_f) \enspace,
\label{impbc}
\end{equation}
where the propagators are taken at the metal-insulator boundary.
Equation (\ref{impbc}) connects the propagator of
an incoming quasiparticle
with Fermi momentum $\bbox{p}_f$ and the propagator of the reflected
quasiparticle with Fermi momentum $\underline{\bbox{p}}_f$
at the interface. For specular reflection the momentum parallel to
the interface is conserved, i.e.,
$\bbox{p}_f^\parallel=\underline{\bbox{p}}_f^\parallel$.
For a complete determination of the quasiclassical propagator we have to
take into account that deep inside the superconductor
the propagator approaches its bulk value.

In realistic systems the reflection of quasiparticles from surfaces and
interfaces is expected to be diffuse, at least to some degree,
due to imperfections at the boundary.\cite{mat,nag96,buch96,burk97}
For isotropic superconductors, elastic scattering of quasiparticles
on walls has no effect on the density of states near the interface,
and thus, on the tunneling current.
The reason for this is that the order parameter for
the reflected quasiparticles is the same for all directions.
In anisotropic superconductors the value of the order parameter
is no longer independent on direction.
A quasiparticle on an outgoing trajectory will
see an order parameter which is, in general, different from
the one on its incoming trajectory. This leads to pair breaking
even for a specularly reflecting wall, and, as a consequence, to
a suppression of the order parameter near the boundary.
For diffuse scattering of quasiparticles at the interface
all outgoing trajectories are mixed.
As a result, sharp structures in the density of states
(e.g., bound states) which would occur for ideal interfaces
are smeared out near rough interfaces.
Thus, the effect of surfaces and interfaces is similar to the
effect of static impurities which leave the order parameter
of isotropic superconductors unaffected (``Anderson's theorem''),
but have destructive influence in anisotropic systems where
superconductivity can even disappear when the impurity concentration
exceeds a critical value.

We model a rough interface by coating an ideal interface
(transparency $D$) on both sides by thin layers
of strongly disordered metals which differ from
the respective bulk materials only by their very short mean free paths
(see Fig.\ 1). The roughness of the interface is measured
by the phenomenological parameter $\rho$ which is essentially
the ratio of the thickness of the layers $\delta$ and
the mean free path $\ell$, i.e., $\rho=\delta/\ell$.
The case $\rho=0$ corresponds to an ideal interface, whereas
$\rho=\infty$ describes a diffuse interface.
The transport of quasiparticles across the disordered layers
is determined by the following transport equation:\cite{cul84}
\begin{equation}
-\left[{\rho\over 2\pi} \big<\hat g^{\: l,r}\big>_{\!\pm},
\hat g^{\: l,r} \right] +i\, \overline{\rm v}_{\perp}^{\: l,r}
\partial_x \hat g^{\: l,r}=0 \enspace .\label{layer}
\end{equation}
The superscripts $l$ and $r$ stand for the left and right sides of the
interface.
The dimensionless velocity $\overline{\rm v}_{\perp}^{\: l,r}$
is the perpendicular to the interface component of the quasiparticle
velocity in the bulk material normalized by an averaged Fermi velocity,
$\overline{\bf v}^{\: l,r}={\bf v}_f^{\: l,r}
/\sqrt{\big<|{\bf v}_f^{\: l,r}|^2\big>_{\bbox{p}_f}}$.
The first term in Eq.\ (\ref{layer}) is Ovchinnikov's anisotropic
scattering self-energy for incoming and outgoing
quasiparticles.\cite{ovc69}
At the ideal interface which separates the two layers
the quasiclassical propagators are discontinuous
and the jump is given by Zaitsev's relations.\cite{zai}
The normal state resistance $R_N$ of the interface turns out to be
independent of the roughness and is given by
\begin{eqnarray}
R_N^{-1}=2e^2 A \int\limits_{{\rm v}_{f\perp}^l>0} {d^2 S^l
\over (2\pi)^3\mid{\bbox{\rm v}}_f^l\mid}
        {\rm v}_{f\perp}^l {D}({\bbox{p}}^l_f) \enspace ,
\label{resist}
\end{eqnarray}
where $A$ is the area of the interface.
A detailed description of this interface model can be found
in Ref.\ \onlinecite{burk97}.

\section{Quasiparticle Bound States at Specularly Reflecting Walls}

In this section, we study in detail the appearance of quasiparticle
bound states near specularly reflecting impenetrable surfaces.
Bound states are localized within a few (or even many; see below)
coherence lengths near the surface, and their energies lie below the
bulk value of the (momentum dependent) gap. Their existence manifests
in $\delta$-peaks in the momentum resolved density of states
(\ref{dos}) below the continuous spectrum.
There are two types of bound states of different origin:
a) zero-energy bound states or midgap states \cite{hu}
   which are a consequence of the change of sign of the order parameter
   along a quasiparticle trajectory which is reflected from the wall,
   and
b) nonzero-energy bound states which emerge due to the depletion of
   the order parameter associated with
   the interface.\cite{buch95,buch81,mat}
While zero-energy bound states are a robust phenomenon of quite
general origin,\cite{ati75}
less information is available for nonzero-energy bound states.

\subsection{Analysis of Quasiparticle Bound States at Specular Walls}

Let the quasiclassical propagator $\hat g$ have a pole at
$\varepsilon=\varepsilon_B(\bbox{p}_f)$,
where
$\varepsilon_B(\bbox{p}_f)$
is the energy of a bound state with Fermi momentum $\bbox{p}_f$
(note that
$\varepsilon_B(\bbox{p} _f)= \varepsilon_B(\underline{\bbox{p}}_f)$).
It is convenient to introduce the residue of the propagator $\hat g$ by
\begin{equation}
\hat{\tilde g}(\bbox{p}_f,\bbox{R};\varepsilon_B(\bbox{p}_f))=
\lim_{\varepsilon\to\varepsilon_B(\bbox{p}_f)}
\left[(\varepsilon-\varepsilon_B(\bbox{p}_f))
\hat g(\bbox{p}_f,\bbox{R};\varepsilon)\right]
\enspace ,
\label{gtilde}
\end{equation}
which is finite
and satisfies the same transport equation as $\hat g$,
but with a modified normalization condition,
\begin{equation}
\big[
\hat{\tilde g}(\bbox{p}_f,\bbox{R};
\varepsilon_B(\bbox{p}_f))
\big]^2=0
\enspace .
\label{mbc}
\end{equation}
The boundary conditions for $\hat{\tilde g}$ coincide with
(\ref{impbc}).
But, in contrast to $\hat g$ the new propagator $\hat{\tilde g}$
vanishes in the bulk of the superconductor.
Thus, the quantity $\hat{\tilde g}$
describes just a quasiparticle state bound to the boundary plane.

For the sake of definiteness, the superconductor is supposed to occupy
the half-space $x>0$ with an impenetrable boundary. Furthermore,
we assume that the gap function $\Delta$ can be chosen real.
Then one obtains from Eilenberger's equations:
\begin{eqnarray}
 \tilde{f}^{+}(\bbox{ p}_{f},x;\varepsilon_B(\bbox{p}_f))=
-\tilde{f}^{*}(\bbox{ p}_{f},x;\varepsilon_B(\bbox{p}_f))
\enspace , \nonumber \\
\tilde g  (\bbox{ p}_{f},x;\varepsilon_B(\bbox{p}_f))=
\tilde g^*(\bbox{ p}_{f},x;\varepsilon_B(\bbox{p}_f))=
|\tilde{f}(\bbox{ p}_{f},x;\varepsilon_B(\bbox{p}_f))|
\enspace .
\label{symmetry}
\end{eqnarray}
The last identity is a consequence of Eq.\ (\ref{mbc}).
We introduce the phase
of the complex anomalous Green function by
\begin{equation}
\tilde{f}(\bbox{p}_f,x;\varepsilon_B(\bbox{p}_f))=
-\tilde{g}(\bbox{p}_f,x;\varepsilon_B(\bbox{p}_f))
\exp\left(i\varphi(\bbox{p}_f,x)\right)
\enspace ,
\label{cp}
\end{equation}
and obtain from (\ref{eil}), (\ref{impbc})
\begin{equation}
\tilde g(\bbox{p}_f,x;\varepsilon_B(\bbox{p}_f))=
\tilde g(\bbox{p}_f,0;\varepsilon_B(\bbox{p}_f))
\exp\left(-\displaystyle\frac{\displaystyle 2}{{\rm v}_{f,x}}
\int\limits^x_0
\Delta(\bbox{p}_f,\tilde x)\sin\left(\varphi(\bbox{p}_f,\tilde x)\right)
d\tilde x\right) \enspace ,
\label{gx}
\end{equation}
together with the following equation for the phase,
\begin{equation}
-\frac{{\rm v}_{f,x}}{2}\partial_x\varphi(\bbox{p}_f,x)+
\varepsilon_B(\bbox{p}_f)-
\Delta(\bbox{p}_f,x)\cos\left(\varphi(\bbox{p}_f,x)\right)=0 \enspace ,
\label{pe}
\end{equation}
and the boundary and asymptotic conditions,
\begin{equation}
\varphi(\bbox{p}_f,0)=\varphi(\underline{\bbox{p}}_f,0) \qquad
\mbox{and} \qquad {\rm v}_{f,x}\Delta_{\infty}(\bbox{p}_f)
\sin\left(\varphi_{\infty}(\bbox{p}_f)\right) >0 \enspace .
\label{bac}
\end{equation}
If $\varphi(\bbox{p}_f,x)$ is a solution of Eqs.\ (\ref{pe}),
(\ref{bac}) with
the energy $\varepsilon_B(\bbox{p}_f)$,
then $\pi-\varphi(\bbox{p}_f,x)$ is a solution with the energy
$-\varepsilon_B(\bbox{p}_f)$.

Since $\partial_x\varphi(\bbox{p}_f,x)$ vanishes in the bulk
($x\rightarrow\infty$),
we immediately get from Eqs.\ (\ref{pe}) and (\ref{bac}) the relation
between the bound state energy $\varepsilon_B(\bbox{p}_f)$ and
the phase in the bulk of the superconductor,
\begin{equation}
\varepsilon_B(\bbox{p}_f)=\Delta_\infty(\bbox{p}_f)
\cos\left(\varphi_{\infty}(\bbox{p}_f)\right)
=\Delta_\infty(\underline{\bbox{p}}_f)
\cos\left(\varphi_{\infty}(\underline{\bbox{p}}_f)\right).
\label{hphi}
\end{equation}
As a consequence, bound states might exist for a given momentum
direction only below the band edges for the momenta $\bbox{p}_f$ and
$\underline{\bbox{p}}_f$, i.e., for
\begin{equation}
|\varepsilon_B(\bbox{p}_f)|\leq
\mathop{\rm min}\nolimits\left\{|\Delta_\infty(\bbox{p}_f)|,
|\Delta_\infty(\underline{\bbox{p}}_f)|\right\} \enspace .
\label{reg}
\end{equation}

Under certain conditions it is possible to find
explicit expressions for the bound state energy
$\varepsilon_B(\bbox{p}_f)$ and the phase $\varphi_{\infty}(\bbox{p}_f)$
in terms of the spatial dependence of the order parameter near the
surface. For this purpose we transform Eq.\ (\ref{pe}) into
the corresponding integral equation,
\begin{eqnarray}
\varphi\left(\bbox{p}_f, x\right)=
\varphi\left(\bbox{p}_f,0\right)+ \frac{\displaystyle
2\Delta_{\infty}(\bbox{p}_f)}{\displaystyle {\rm v}_{f,x}}
\int\limits^{x}_0\Biggl(
\cos\left(\varphi_\infty\left(\bbox{p}_f\right)\right) -
\qquad\qquad\qquad\qquad\qquad \nonumber \\
\qquad\qquad\qquad\qquad\qquad\qquad\qquad\qquad\qquad
-\frac{\displaystyle
\Delta\left(\bbox{p}_f,\tilde x\right)}{\displaystyle
\Delta_\infty(\bbox{p}_f)}\cos\left(\varphi\left(\bbox{p}_f,
\tilde x\right)\right)\Biggr) d\tilde x \enspace .
\label{ipe}
\end{eqnarray}
Furthermore, we suppose that
the phase $\varphi(\bbox{p}_f,x)$ deviates only
weakly from its value at the surface everywhere within the
superconductor
($|\varphi(\bbox{p}_f,x)-\varphi(\bbox{p}_f,0)|\ll 1$).
Then, as a first approximation, one can substitute
on the right hand side of Eq.\ (\ref{ipe}),
$\cos\left(\varphi\left(\bbox{p}_f,x\right)\right)
\rightarrow \cos\left(\varphi_{\infty}\left(\bbox{p}_f\right)\right)-
\sin\left(\varphi_{\infty}\left(\bbox{p}_f\right)\right)
\left(\varphi\left
(x,\bbox{p}_f\right)-\varphi_{\infty}\left(\bbox{p}_f\right)\right)$,
which leads to
\begin{equation}
\varphi_\infty\left(\bbox{p}_f\right)=
\varphi\left(\bbox{p}_f,0\right)+ \frac{\displaystyle
\Delta_{\infty}(\bbox{p}_f)}{\displaystyle \Delta_{max}}
\cos\left(\varphi_\infty\left(\bbox{p}_f\right)\right)A(\bbox{p}_f,
\varphi_{\infty}\left(\bbox{p}_f\right)) \enspace ,
\label{iep}
\end{equation}
where
\begin{equation}
A(\bbox{p}_f,\varphi_{\infty}\left(\bbox{p}_f\right))=
\frac{\displaystyle 2\Delta_{max}}{\displaystyle {\rm v}_{f,x}}
\int\limits^\infty_0
\exp\left(-\frac{\displaystyle 2}{\displaystyle {\rm v}_{f,x}}
\sin\left(\varphi_{\infty}\left(\bbox{p}_f\right)\right)
\int\limits^{x}_0 \Delta\left(\bbox{p}_f,
\tilde x \right)
d\tilde x\right)
\Biggl(1 -\frac{ \displaystyle \Delta\left(\bbox{p}_f,x \right)}
{\displaystyle \Delta_\infty(\bbox{p}_f)}\Biggr) dx.
\label{A}
\end{equation}
Here, we introduced the quantity
$\Delta_{max}=
\mathop{\rm max}\limits_{S_F}
\left\{|\Delta_\infty(\bbox{p}_f)|\right\}$.

The condition of weak deviations of the phase from its value on the
surface is transformed with the help of Eqs.\ (\ref{iep}), (\ref{A})
into the form
\begin{equation}
\frac{\displaystyle
|\Delta_{\infty}(\bbox{p}_f)|}{\displaystyle \Delta_{max}}\,
|\cos\left(\varphi_\infty\left(\bbox{p}_f\right)\right)|\,
|A(\bbox{p}_f,\varphi_{\infty}\left(\bbox{p}_f\right))|\ll 1 \enspace .
\label{cond}
\end{equation}
Thus, the approximation can be justified, for example,
for momentum directions for which
the order parameter is only slightly suppressed at the surface
(when
$|A(\bbox{p}_f,\varphi_{\infty}\left(\bbox{p}_f\right))|$
is small), or in the vicinity of
nodes of the order parameter (when
$|\Delta_{\infty}(\bbox{p}_f)|/ \Delta_{max}$ is small ),
or for a small enough
$|\cos\left(\varphi_\infty\left(\bbox{p}_f\right) \right)|$.

If condition (\ref{cond}) is valid both for incoming and for outgoing
momenta, we get from Eqs.\ (\ref{bac}), (\ref{hphi}), (\ref{iep})
the following equation for the bound state energy,
\begin{eqnarray}
\frac{\displaystyle
\varepsilon_B^2(\bbox{p}_f)}{\displaystyle \Delta_\infty(\bbox{p}
_f)\Delta_\infty (\underline{\bbox{p}}_f)} -\mathop{\rm sgn}\nolimits
\left(\Delta_\infty(\bbox{p}_f) \Delta_\infty (\underline{\bbox{p}}
_f)\right)\left(1- \frac{\displaystyle
\varepsilon_B^2(\bbox{p}_f)}{\displaystyle
\Delta^2_\infty(\bbox{p}_f)}\right)^{1/2}\left(1- \frac{\displaystyle
\varepsilon_B^2(\bbox{p}_f)}{\displaystyle
\Delta^2_\infty(\underline{\bbox{p}}_f)}\right)^{1/2}= \nonumber \\
\qquad\qquad\qquad\qquad\qquad
=\cos\Biggl(\frac{\displaystyle \varepsilon_B(\bbox{p}_f)}{\displaystyle
\Delta_{max}}\left[A(\bbox{p}_f,\varphi_{\infty}\left(\bbox{p}_f\right))
-A(\underline{\bbox{p}}_f,\varphi_{\infty}(\underline{\bbox{p}}_f)
)\right]\Biggr)
\enspace .
\label{eq}
\end{eqnarray}
It is easy to see that the condition
$\Delta_\infty(\bbox{p}_f) \Delta_\infty (\underline{\bbox{p}}_f)<0$
implies the existence of a midgap state at
$\varepsilon_B(\bbox{p}_f)=0$.
We should note that this solution does not depend on the validity
of our expansion (\ref{iep}) for the phase
$\varphi_\infty\left(\bbox{p}_f\right)$.
If the order parameter changes sign along a trajectory, then
$\varphi(\bbox{p}_f)=(\pi/2)\mathop{\rm sgn}\nolimits
\left({\rm v}_{f,x}\Delta_\infty(\bbox{p}_f)\right)$ is a solution
of Eqs.\ (\ref{pe}), (\ref{bac}) for $\varepsilon_B(\bbox{p}_f)=0$.
The left-hand side of Eq.\ (\ref{cond}) is equal to zero for this
solution.
Another consequence of Eq.\ (\ref{eq}) is the absence of
nonzero-energy states near the edge of the continuous
spectrum (that is close to the right-hand side of (\ref{reg}))
provided
$\Delta_\infty(\bbox{p}_f) \Delta_\infty (\underline{\bbox{p}}_f)<0$
and $|A(\bbox{p}_f,\varphi_{\infty}\left(\bbox{p}_f\right))
\Delta_{\infty}(\bbox{p}_f)|/ \Delta_{max}\ll 1$.
However, bound states may appear near
the band edge if the parameter
$|A(\bbox{p}_f,\varphi_{\infty}\left(\bbox{p}_f\right))
\Delta_{\infty}(\bbox{p}_f)|/ \Delta_{max}$ is not small
(see Fig.\ 4c).
In this case, our approximation (\ref{iep}) for the phase
$\varphi_\infty\left(\bbox{p}_f\right)$ breaks down.

Let us now consider the opposite case,
$\Delta_\infty(\bbox{p}_f) \Delta_\infty (\underline{\bbox{p}}_f)>0$.
If the condition $ |\Delta_{\infty}(\bbox{p}_f)
A(\bbox{p}_f,\varphi_{\infty}\left(\bbox{p}_f\right))|/\Delta_{max}
\ll 1$ is satisfied for a pair of incoming and outgoing momenta,
we see from Eqs.\ (\ref{reg}) and (\ref{eq}) that bound states occur
only for trajectories with
$|\Delta_{\infty}(\bbox{p}_f)-
\Delta_{\infty}(\underline{\bbox{p}}_f)| \ll
|\Delta_{\infty}(\bbox{p}_f)|$.
The energies of these bound states lie close to the continuum.
These bound states arise for momentum directions
close to the normal or almost parallel to the boundary plane
where $\Delta_{\infty}(\bbox{p}_f) \approx
\Delta_\infty (\underline{\bbox{p}}_f)$.
The latter region of momentum
directions, however, is not so interesting because of its comparatively
small contribution to the tunneling current.
For bound state energies near the edge of the continuum
($\varepsilon_B(\bbox{p}_f) \approx \Delta_{\infty}(\bbox{p}_f)$)
we find from Eq.\ (\ref{hphi})
$|\sin\left(\varphi_\infty(\bbox{p}_f)\right)| \ll 1$.
Therefore, we can approximate Eq.\ (\ref{A}) by
\begin{equation}
A_0(\bbox{p}_f)=A(\bbox{p}_f,0)=
\frac{\displaystyle 2\Delta_{max}}{\displaystyle {\rm v}_{f,x}}
\int\limits^\infty_0
\Biggl(1 -\frac{ \displaystyle \Delta\left(\bbox{p}_f,x \right)}
{\displaystyle \Delta_\infty(\bbox{p}_f)}\Biggr) dx \enspace ,
\label{an}
\end{equation}
and
obtain the following expression for bound states energies near the
band edge:
\begin{eqnarray}
\frac{\displaystyle \varepsilon_B^2(\bbox{p}_f^2)}{\displaystyle
\Delta _{max}^2}=
\frac{\displaystyle\Delta _\infty (\bbox{p}_f)\Delta _\infty (
\underline{\bbox{p}}_f)}{\displaystyle\Delta _{max}^2}\left( 1-\frac{
\displaystyle\Delta _\infty (\bbox{p}_f)
\Delta _\infty (\underline{\bbox{p}}_f)}{4\Delta _{max}^2}
\left( A_0(\bbox{p}_f)-A_0(\underline{\bbox{p}}_f)\right) ^2\right) +
\quad \quad \qquad \nonumber \\
+\left( \sqrt{\frac{\displaystyle
\Delta _\infty (\bbox{p}_f)}{\displaystyle
\Delta _\infty (\underline{\bbox{p}}_f)}}-\sqrt{\frac{\displaystyle
\Delta _\infty (\underline{\bbox{p}}_f)}{\displaystyle
\Delta _\infty (\bbox{p}_f)}}\right) ^2\left( \frac{\displaystyle
\Delta _\infty (\bbox{p}_f)\Delta _\infty (\underline{\bbox{p}}_f)}{
\displaystyle
4\Delta _{max}^2}-\frac{\displaystyle 1}{\displaystyle
\left( A_0(\bbox{p}_f)-A_0(\underline{\bbox{p}}_f)\right)^2}\right)
\enspace .
\label{epsp}
\end{eqnarray}

It is of particular interest to consider the momentum direction along
the normal to the boundary plane
$\bbox{p}_f^n\|\bbox{\hat n}$,
where the
additional relation $\underline{\bbox{p}}_f^n=-\bbox{p}_f^n$ holds.
Using $\Delta(\bbox{p}_f)=\Delta(-\bbox{p}_f)$
we get
$\varphi(\bbox{p}_f^n,x)=-\varphi(\underline{\bbox{p}}_f^n,x)$.
Together with the boundary condition (\ref{bac})
for $\varphi$ at the wall
we find $\varphi(\bbox{p}_f^n,0)=0$.
Then one has
$|\varphi_\infty(\bbox{p}_f^n)-\varphi(\bbox{p}_f^n,0)| =
|\varphi_\infty(\bbox{p}_f^n)| \ll 1$, and obtains from
Eqs.\ (\ref{hphi}), (\ref{iep})
the following expression for the bound state energy:
\begin{equation}
|\varepsilon_B(\bbox{p}_f^n)|=|\Delta_\infty(\bbox{p}_f^n)|
\cos\left(\frac{
\Delta_\infty(\bbox{p}_f^n)}{\Delta_{max}}A_0(\bbox{p}_f^n)\right)
\approx \left(1-
\frac{\Delta^2_\infty(\bbox{p}_f^n)}{2\Delta^2_{max}} A_0^2(\bbox{p}
_f^n)\right)|\Delta_\infty(\bbox{p}_f^n)| \enspace .
\label{bse}
\end{equation}
The asymptotic condition (\ref{bac}) for $\varphi$ is satisfied
for this state for positive values of $A_0(\bbox{p}_f^n)$, hence
\begin{equation}
\int\limits^\infty_0
\left(
1 -\frac{\displaystyle
\Delta\left(\bbox{p}_f^n,x\right)}{\displaystyle
\Delta_\infty(\bbox{p}_f^n)}
\right)
dx>0 \enspace .
\label{acn}
\end{equation}

This inequality is fulfilled, in general, for the order parameter with
the amplitude near the wall being less than its bulk value. This is
true, in particular, for a one-component order parameter.
Condition (\ref{acn}) might be violated
only if order parameter components with different symmetries appear
in the vicinity of the surface
due to nonzero coupling constants in subdominant symmetry channels.
\cite{buch95,mat}
In the absence of this effect, the function
$\Delta\left(\bbox{p}_f^n, x\right)$
is always a monotonously increasing function of the distance
from the boundary, which ensures the validity of Eq.\ (\ref{acn}).
Thus, we have shown in Eq.\ (\ref{bse})
that bound states with nonzero energy do exist for
$\bbox{p}_f\|\bbox{\hat n}$ for any (even tiny) suppression of the order
parameter
at the surface. If one interprets these states as bound states in the
effective potential well formed by the order parameter at the boundary,
one can say that for $\bbox{p}_f\|\bbox{\hat n}$ quasiparticle bound
states appear in any shallow potential well.

One can find the
dependence of the bound state energy on the momentum direction
in the vicinity of perpendicular incidence,
i.e., for $\bbox{p}_f \approx
\bbox{p}_f^n$.
Assuming that the boundary is a symmetry plane of the Fermi surface
of the superconductor, one may write
\begin{equation}
\bbox{p}_f=(\bbox{p}_f \hspace*{-2pt}\cdot\hspace*{-2pt}
\bbox{\hat n})\bbox{\hat n}+
\delta \bbox{p}_f\equiv \bbox{p}_f^n+\delta \bbox{p}_f,
\qquad
\underline{\bbox{p}}_f=\delta \bbox{p}_f-\bbox{p}_f^n
 \enspace.
\end{equation}
Furthermore, we expand  all the quantities on the right-hand side of
Eq.\ (\ref{epsp}) in powers of small deviations $\delta \bbox{p}_f$.
In particular,
\begin{equation}
\Delta _\infty (\bbox{p}_f)=\Delta _\infty (\bbox{p}_f^n)+
\left( \delta \bbox{p}_f \hspace*{-2pt}\cdot\hspace*{-2pt}
\nabla _{\bbox{p}_f}\right)
\Delta _\infty (\bbox{p}_f^n)+\frac 1
2\left( \delta\bbox{p}_f
\hspace*{-2pt}\cdot\hspace*{-2pt} \nabla _{\bbox{p}_f}\right)^2
\Delta _\infty (
\bbox{p}_f^n)+\ldots \enspace.
\end{equation}
By keeping only terms up to second order in $\delta \bbox{p}_f$,
expression (\ref{epsp}) for the bound state energy can be reduced
to
\begin{eqnarray}
\frac{\varepsilon_B^2(\bbox{p}_f)}{\Delta _{max}^2}
=\frac{\Delta _\infty ^2(\bbox{p}_f^n)}{
\Delta _{max}^2}\left( 1-A_0^2(\bbox{p}_f^n)
\frac{\Delta _\infty ^2(\bbox{p}_f^n)}{\Delta _{max}^2}\right) -
\frac{\Delta _\infty ^4(\bbox{p}_f^n)}{\Delta _{max}^4}A_0(
\bbox{p}_f^n)\left(
\delta \bbox{p}_f\hspace*{-2pt}\cdot\hspace*{-2pt}
\nabla _{\bbox{p}_f} \right)^2 A_0(\bbox{p}_f^n)-
\nonumber \\
-\frac{\left( \left(
\delta\bbox{p}_f \hspace*{-2pt}\cdot\hspace*{-2pt}
\nabla _{\bbox{p}_f} \right)
\Delta _\infty (\bbox{p}_f^n)\right) ^2}{\Delta _\infty^2(\bbox{p}_f^n)}
\left( \frac 1{A_0^2(\bbox{p}_f^n)}-
2\frac{\Delta _\infty ^4(\bbox{p}_f^n)}{\Delta
_{max}^4}A_0^2(\bbox{p}_f^n)\right)+
\qquad \nonumber \\
+\frac{\left( \delta\bbox{p}_f \hspace*{-2pt}\cdot\hspace*{-2pt}
\nabla _{\bbox{p}_f} \right) ^2
\Delta _\infty (\bbox{p}_f^n)}{\Delta _\infty (
\bbox{p}_f^n)}
\left( \frac{\Delta _\infty ^2(\bbox{p}_f^n)}{\Delta _{max}^2}-2
\frac{\Delta _\infty ^4(\bbox{p}_f^n)}
{\Delta _{max}^4}A_0^2(\bbox{p}_f^n)\right) +\ldots
\enspace .
\label{epd}
\end{eqnarray}
The linear terms in $\delta \bbox{p}_f$ cancel
each other, and hence the bound state energy has a local extremum
for the direction normal to the boundary.

It follows from condition (\ref{reg}), that bound states
occur only in a narrow region of the Fermi surface
for momentum directions which are close to the normal of the interface.
Indeed, expanding Eq.\ (\ref{reg}) in the vicinity of $\bbox{p}_f^n$,
\begin{equation}
\min \left\{ |\Delta _\infty ^2(\bbox{p}_f)|,|\Delta _\infty ^2
(\underline{ \bbox{p}}_f)|\right\} =\Delta _\infty ^2
(\bbox{p}_f^n)-2\left| \Delta _\infty (\bbox{p}_f^n)
\left( \delta\bbox{p}_f \hspace*{-2pt}\cdot\hspace*{-2pt}
\nabla _{\bbox{p}_f} \right)
\Delta _\infty (\bbox{p}_f^n)\right| +\ldots \enspace ,
\label{min}
\end{equation}
and keeping only terms up to second order in $\delta \bbox{p}_f$,
we get from (\ref{reg}), (\ref{epd}), (\ref{min}) the following
range for the occurance of bound states:
\begin{equation}
\left| \frac{\left(
\delta\bbox{p}_f \hspace*{-2pt}\cdot\hspace*{-2pt}
\nabla _{\bbox{p}_f^n} \right) \Delta _\infty (\bbox{p}_f^n)}
{\Delta _\infty (\bbox{p}_f^n)}\right| <\frac 58\frac{
\Delta _\infty ^2(\bbox{p}_f^n)}{\Delta _{max}^2}A_0^2(\bbox{p}_f^n)
\enspace .
\label{regi}
\end{equation}
One can see from (\ref{cond}) and condition
$|\varphi_\infty(\bbox{p}_f^n)|\ll 1$,
that the right-hand side of Eq.\ (\ref{regi}) is much smaller than 1.

\subsection{Results for d-Wave Superconductors}

Now, we consider a tetragonal
$d_{x^2-y^2}$
superconductor
with a cylindrical Fermi surface whose
$z_0$-axis
is parallel to the
boundary plane. The order parameter is supposed to have the form
\begin{equation}
\Delta \left( \bbox{p}_f,x\right) =\Delta (x)\cos (2\phi -2\alpha)
\enspace ,
\label{gap}
\end{equation}
where $\phi$ is
the azimuth angle between the Fermi momentum $\bbox{p}_f$
and the surface normal
$\bbox{\hat n}$  (which is taken as the $x$ direction).
The angle $\alpha$
describes the orientation of the crystalline axis $\bbox{\hat x_0}$
with respect to the normal of the boundary.
For this
case, Eq.\ (\ref{epd}) for the bound state energy takes the form:
\begin{eqnarray}
\frac{\varepsilon_B^2(\phi)}{\Delta_{max}^2\cos^2(2\alpha)}=
1-A_0^2\cos ^2(2\alpha)-
\qquad\qquad\qquad\qquad \qquad \qquad \qquad \qquad \qquad
\nonumber \\
\qquad\qquad\qquad\qquad\qquad\qquad
\phi ^2\left(\!\! A_0^2\cos ^2(2\alpha)+4(1-2A_0^2)+
4\frac{ \tan ^2(2\alpha)}{A_0^2\cos ^2(2\alpha)}\right) \enspace .
\label{edw}
\end{eqnarray}
Consequently, the angular range
for which  bound states exist is limited by
\begin{equation}
|\phi |<\frac{5A_0^2\cos ^2(2\alpha)}{16\tan (2\alpha)} \enspace ,
\enspace\enspace
0\leq \alpha\leq \frac \pi 4\enspace .
\label{condw}
\end{equation}
The quantity $A_0=A_0(\phi=0)$
depends on the misorientation angle $\alpha$ as well as on the
temperature $T$.
According to (\ref{condw}), bound states for quasiparticles
with momentum directions close to the normal of the boundary
always appear,
with the exception of two obvious
crystalline orientations:
a) $\alpha= 0$ where the order parameter is not suppressed at the
interface ($A_0$ vanishes for $\alpha \rightarrow 0$), and b)
$\alpha=\pm 45^\circ$ where the order parameter for perpendicular
incidence is zero.

According to (\ref{gx}), (\ref{gap}) the characteristic
length describing the localization of bound states near the surface is
of the order of
${\rm v}_{f}\cos\phi/\left(\Delta _\infty \left(\bbox{p}_f\right)
\sin\left(\varphi _\infty \left(\bbox{p}_f\right) \right) \right)
\sim \xi(T)\cos\phi/\sin\left(\varphi_\infty\right)\cos(2\phi-2\alpha)$.
For momentum directions, for which the bound levels lie close
to the edge of the continuous spectrum, one gets from (\ref{hphi})
$|\sin\left(\varphi _\infty(\bbox{p}_f)\right)| \ll 1$.
The characteristic length is in this case much greater than
the temperature dependent coherence length $\xi(T)$
(provided $\cos\phi$ is not very small).
The factor $\cos(2\phi-2\alpha)$ in the denominator may result
in an additional increase of the characteristic length.

The results of numerical calculations for the angular dependence of
the bound state energy $\varepsilon_B(\phi)$
are presented in Fig.\ 2 for various misorientation angles $\alpha$.
The temperature is $T=0.1\,T_c$.
The spatially varying gap amplitude $\Delta (x)$
was evaluated self-consistently (see Ref.\ \onlinecite{buch95}).
Bound states with nonzero energy
$\varepsilon_B(\phi)$
occur in a comparatively narrow angular
region in the vicinity of the surface normal (Fig.\ 2a).
These bound states are also described analytically above.
Bound states which occur for
lattice to surface orientations $\alpha\gtrsim 37^\circ$ (Fig.\ 2b)
(it is sufficient to consider $0^\circ\le\alpha\le 45^\circ$)
correspond to values of the parameter
$|\Delta_{\infty}(\bbox{p}_f)
\cos\left(\varphi_\infty\left(\bbox{p}_f\right)\right)
A_0(\bbox{p}_f,\varphi_{\infty}\left(\bbox{p}_f\right))|/\Delta_{max}$
of the order of unity for which condition (\ref{cond}) is violated.
They arise for trajectories for which
$\Delta_\infty(\bbox{p}_f) \Delta_\infty (\underline{\bbox{p}}_f)<0$.
In general, there are also bound states
for glancing angles, i.e., $\phi \approx \pm 90^\circ$,
but they give only a very small contribution to the tunneling  current,
and we do not plot them here.
The limit for bound states with nonzero quasiparticle energy
is determined by momentum directions for which discrete bound state
levels merge in the continuum.
Note that for misorientation angles
$37^\circ \lesssim \alpha < 45^\circ$
bound states appear for $\varepsilon=0$ (midgap states), for
$\varepsilon \approx \Delta_\infty (\phi=0^\circ)=
                     \Delta_{max}\cos(2\alpha)$
(perpendicular incidence) and for
$\varepsilon \approx \Delta_\infty (\phi=45^\circ)=
                     \Delta_{max}\sin(2\alpha)$
($\phi=45^\circ$).
As a result one finds three peaks in the conductance
(see inset of Fig.\ 6a).

In Fig.\ 3
we display the dependence of the bound state energy
for perpendicular incidence ($\phi =0$)
on the misorientation angle $\alpha$
for $T=0$.
We included the function
\mbox{$\Delta _\infty (\phi =0)/\Delta_{max}=\cos(2\alpha)$}
to demonstrate that the bound states with nonzero energy
lie very close to the band edge.

The angle-resolved density of states $\nu(\phi,x=0;\varepsilon)$
at clean surfaces ($\rho=0$) is shown in Fig.\ 4 for the
misorientation angles $\alpha=0^\circ$, $20^\circ$ and $45^\circ$.
For a (100) surface ($\alpha=0^\circ$) (Fig.\ 4a)
the order parameter is homogeneous up to the wall,
and the surface density of states coincides with its bulk value.
In particular, there are no surface bound states.
The divergences presented in Fig.\ 4a lie at the edges of the
continuum ($\varepsilon=|\Delta_\infty(\phi)|$). For a lattice to
surface orientation $\alpha=20^\circ$ (Fig.\ 4b) bound states
occur near the surface. For perpendicular incidence ($\phi=0^\circ$)
there is a bound state with nonzero energy near the band edge,
while for $\phi=30$ and $45^\circ$ midgap states appear
(cf.\ Refs.\ \onlinecite{mat,buch95}).
In Fig.\ 4c we display the surface density of states for an ideal
(110) surface ($\alpha=45^\circ$).
For a quasiparticle moving along a trajectory with angle $\phi=45^\circ$
we find both a midgap state and a bound state
near the edge of the continuum.
Both arise for an ideal boundary under the condition
$\Delta _\infty (\bbox{p}_f)\Delta _\infty (\underline{\bbox{p}}_f)<0$.
For this special surface orientation we have
$\Delta_\infty(\bbox{p}_f)=-\Delta_\infty (\underline{\bbox{p}}_f)$
for all pairs of incoming and reflected trajectories.
From Eqs.\ (\ref{pe}) and (\ref{bac}) we find
$\varphi(\bbox{p}_f,x)=\pi-\varphi(\underline{\bbox{p}}_f,x)$,
and at the surface
$\varphi(\bbox{p}_f,0)=\varphi(\underline{\bbox{p}}_f,0)=\pi/2$.
On the other hand, we obtain from (\ref{hphi})
for bound states near the continuum
$\varphi_\infty(\bbox{p}_f)=\pi-\varphi_\infty(\underline{\bbox{p}}_f)
\approx 0$ or $\pi$.
Hence, the phase $\varphi$ has to vary comparatively rapidly with
spatial position $x$. For this reason,
our analytical approach, based on the approximation
(\ref{iep})-(\ref{cond}), breaks down for this special case.

The influence of surface roughness is shown in Fig.\ 5
for a lattice to surface orientation $\alpha=20^\circ$.
We observe the broadening of the bound states with increasing roughness.
The midgap state is smeared out faster than
the nonzero-energy bound state (Figs.\ 5a, b). The point is that
interference effects at the surface,
which are of importance for the midgap states,
are destroyed by diffuse scattering of quasiparticles
at the boundary.
On the other hand, for the existence of the nonzero-energy bound state
only the suppression of the order parameter near the wall
(even if it is very small) plays a role.
For very rough surfaces (Fig.\ 5c) the midgap state is completely
suppressed whereas the nonzero-energy bound state merges in the
continuum.
A similar behavior of midgap states
at rough surfaces was observed in Ref.\nolinebreak\onlinecite{mat}.

\section{Temperature Dependence of the Peaks in the
	 Differential Conductance}

We consider a tunnel junction with transparency ${D} \ll 1$.
First, we focus on specularly reflecting interfaces
between a normal metal and a clean singlet anisotropically
paired superconductor. Let the normal of
the junction barrier plane be directed along the $x$-axis
($\bbox{\hat n}\parallel \bbox{\hat x}$). The superconductor occupies
the left half-space ($x<0$, index $l$) and the normal metal the right
one ($x>0$, index $r$).
Then,
the dissipative current in first order
in the barrier transparency $D$ may be represented by
\cite{bs1}
\begin{equation}
I_x(V)=-eA\!\int\limits_{{\rm v}_{f,x}^l>0}\frac{d^{\,2}S^l}{(2\pi )^3}
\frac{{\rm v}_{f,x}^l}{|\bbox{\rm v}_{f}^l|}{D}
\left(\bbox{p}_{f}^l\right)
\int\limits_{-\infty }^\infty
d\varepsilon \left(
\tanh \left( \frac{\varepsilon -eV}{2T}\right)-
\tanh \left( \frac\varepsilon {2T}\right) \right)
\nu_S\left( \bbox{ p}_{f}^l,0;\varepsilon \right)
\enspace .
\label{j}
\end{equation}
Here, $\nu_S(\bbox{p}_f^{l},x=0;\varepsilon)$
is the angle resolved density of states (\ref{dos})
of the superconductor on the left side of the tunneling barrier.
It is calculated for zero transparency ($D=0$).
The voltage $V=\Phi^l - \Phi^r$ is the difference between the electric
potentials of the left and right electrodes,
and $A$ is the area of the interface.
Properties of the normal metal enter only implicitly via the
function $D(\bbox{p}_f)$, which is the probability of a
quasiparticle with momentum $\bbox{p}_f$ to tunnel across the barrier.
Note that the charge $e$ is negative in our notation.

Most experiments measure directly the differential conductance
$G=dI/dV$. From Eq.\ (\ref{j}) we get
\begin{equation}
G(V,T)=\frac{e^2 A}{2T}\!\int\limits_{{\rm v}_{f,x}^l>0}
\frac{d^{\,2}S^l}{(2\pi )^3}
\frac{{\rm v}_{f,x}^l}{|\bbox{\rm v}_{f}^l|}{D}
\left(\bbox{p}_{f}^l\right)
\int\limits_{-\infty }^\infty d\varepsilon \,
\frac{\nu_S( \bbox{ p}_{f}^l,0;\varepsilon)}
{\cosh^2 \left( \frac{\varepsilon -eV}{2T}\right)}
\enspace ,
\label{g}
\end{equation}
which reduces at zero temperature to
\begin{equation}
G(V,T=0)=2e^2 A\!\int\limits_{{\rm v}_{f,x}^l>0}
\frac{d^{\,2}S^l}{(2\pi )^3}
\frac{{\rm v}_{f,x}^l}{|\bbox{\rm v}_{f}^l|}{D}
\left(\bbox{p}_{f}^l\right)
\nu_S( \bbox{ p}_{f}^l,0;eV)
\label{g0}
\enspace .
\end{equation}
Thus, tunneling experiments at low temperatures measure
essentially the surface density of states
integrated over the Fermi surface.
As we will show below the surface density of states of anisotropic
superconductors deviates substantially from its bulk value
if the surface is pair breaking.
Hence, tunneling experiments are extremely sensitive
to special features of the surface (orientation and roughness).

We notice, that in case of a real order parameter
the density of states is symmetric with respect to the Fermi
energy $\varepsilon_f$, i.e.,
$\nu(\bbox{p}_{f},\bbox{R};\varepsilon)=
\nu(\bbox{p}_{f},\bbox{R};-\varepsilon)$.
Consequently, the conductance in Eq.\ (\ref{g}) is symmetric:
$G(V)=G(-V)$.
Hence, the asymmetry of the conductance observed experimentally
in Ref.\ \onlinecite{renfisch} is beyond the scope of this article.

\subsection{Surfaces without Pair Breaking}

Let us first consider
surfaces which are not
pair breaking.
Then,
the density of states at the
interface
coincides with its bulk value,
\begin{equation}
\nu_S\left( \bbox{ p}_{f },0;\varepsilon \right)=\frac{|\varepsilon |}
{\sqrt{\varepsilon ^2-|\Delta (\bbox{ p}_{f })|^2}}
\,\Theta\left(|\varepsilon| -|\Delta (\bbox{ p}_{f })| \right)
\enspace .
\label{gbulk}
\end{equation}
At low temperatures ($T\ll |eV|$) we get
the following expression for the
differential conductance
($\Delta_0(\bbox{ p}_{f })=\Delta(\bbox{ p}_{f },T=0)$):
\begin{equation}
G\approx e^2 A\int\limits_{{\rm v}_{f,x}^l>0}\frac{d^{\,2}S^l}{(2\pi)^3}
\frac{{\rm v}_{f,x}^l}{|\bbox{\rm v}_{f}^l|}{D}
\left(\bbox{ p}_{f }^l\right)
\int\limits_{-\infty }^\infty
\frac{d\tilde{\varepsilon}}{\cosh^2\left(\tilde{\varepsilon}\right)}
\frac{|eV|
\Theta\left(|eV|-|\Delta_0(\bbox{ p}_{f })|+
2T\tilde{\varepsilon}\mathop{\rm sgn}\nolimits(eV)\right)}
{\sqrt{4eVT\tilde{\varepsilon} +
\left((eV)^2-|\Delta_0(\bbox{ p}_{f })|^2\right)}}
\enspace .
\label{Gbulk}
\end{equation}

In the case of an isotropic $s$-wave superconductor
we obtain for $T\ll||eV|-\Delta_0|$
the well-known result for the conductance,
\begin{equation}
G=\frac{1}{R_N}\frac{|eV|}
{\sqrt{(eV)^2-
\Delta^2_0}}
\,\Theta\left(|eV|-\Delta_0\right)
\enspace .
\label{Gi1}
\end{equation}
At zero temperature
the conductance diverges in the limit $|eV|\rightarrow\Delta_0$,
which is directly associated with the singularity in the BCS density
of states at $\varepsilon=\Delta_0$.
It is worth noting that in the opposite limiting case,
$||eV|-\Delta_0|\ll T\ll \Delta_0$, we obtain instead of (\ref{Gi1})
the following low-temperature anomaly in the conductance:
\begin{equation}
G
\approx
\frac{0.48}{R_N}\sqrt{\frac{\Delta_0}{T}}
\enspace .
\label{Gi2}
\end{equation}
Hence, for the voltage $|eV|=\Delta_0$ there is a divergence of the
conductance, $G\propto 1/\sqrt{T}$, in the low-temperature limit,
due to the singularity in the BCS density of states.

It has been known for a long time
that anisotropy of the superconducting order parameter
washes out structures in the integrated density of states,
and correspondingly removes (or at least reduces)
the singular behavior of the conductance (see Eq.\ (\ref{g0})).
This statement is valid, in general,
in the absence of pair breaking at the surface.
At the same time, in the presence of nodes of the order parameter on the
Fermi surface, a nonzero subgap conductance appears. In particular,
specific gapless and, in general, non-Ohmic behavior takes place in the
limit of small voltages and low temperatures.
For three-dimensional $d$-wave superconductor the
singular behavior of the conductance vanishes in the absence of surface
pair breaking. For the low voltage behavior one has, depending
on the crystal orientation of such a superconductor relative to
the barrier plane:\cite{bgz}
$G\propto |eV|/(R_N\Delta_0)$ or $G\propto
(|eV|/R_N\Delta_0)\ln(\Delta_0/|eV|)$ under the condition
$T\ll |eV|\ll \Delta_0$, and $G\propto (T/R_N\Delta_0)$ or
$G\propto (T/R_N\Delta_0)\ln(\Delta_0/T)$ if $|eV|\ll T\ll \Delta_0$.

Now, we consider a quasi two-dimensional Fermi surface, e.g.,
a Fermi cylinder with its principal axis parallel to the barrier plane.
Let $\bbox{s}=(s_\|,s_\perp)$ be a two-dimensional parameter set
which describes the Fermi surface,
i.e., $\bbox{p}_f=\bbox{p}_f(\bbox{s})$.
As all quantities depend only on $s_\|$
the order parameter
can be approximated in the vicinity of a local maximum by
$|\Delta (\bbox{p}_f)|=\Delta_{max}(1-bs_\|^2)$.
Substituting this expansion into (\ref{Gbulk}) we find the following
singular contribution to the conductance (coming from the region
$bs_\|^2\ll 1$):
\begin{equation}
G_s\approx -e^2 A\,\sqrt{\frac 2b}\int \frac{ds_\perp }{(2\pi )^3}
\left|\frac{\partial \bbox{p}_f^l}{\partial s_{\|}}\right|
\left|\frac{\partial \bbox{p}_f^l}{\partial s_{\perp}}\right|
\frac{{\rm v}_{f,x}^l}{|\bbox{\rm v}_{f}^l|}
{D} (\bbox{p}_f^l)
\left\{\enspace \begin{array}{l}
\ln \left|\frac{\displaystyle |eV|}{\displaystyle\Delta _0}-1\right|,
\quad T\ll ||eV|-\Delta _0|, \enspace \Delta _0 \enspace ,\\
\frac{\displaystyle 1}{\displaystyle 2}
\ln \left(\frac{\displaystyle T}{\displaystyle \Delta _0}\right),
\quad ||eV|-\Delta _0|\ll T\ll \Delta_0 \enspace ,
\end{array}
\right.
\label{Gsing}
\end{equation}
where $\Delta_0=\Delta_{max}(T=0)$.
For the low voltage behavior one finds in this case in the presence of
nodes of the order parameter
$G\propto |eV|/R_N\Delta_0$ for $T\ll |eV|\ll \Delta_0$ and
$G\propto T/(R_N\Delta_0)$ for $|eV|\ll T\ll \Delta_0$.

So, in case of a $d$-wave superconductor
with a cylindrical Fermi surface the low-temperature anomaly
in the conductance as well as the singular dependence on the voltage
turns out to
reduce to a logarithmic singularity,
$G \propto \ln (T)\, ,\ln (||eV|-\Delta _0|)$,
which is much weaker compared to
the $s$-wave case where we found
$G\propto T^{-1/2}, \,
\left((eV)^2-\Delta^2_0\right)^{-1/2}$.

\subsection{Surfaces with Pair Breaking}

For ideally smooth surfaces with pair breaking
the singular behavior of the conductance of an NIS junction
with a $d$-wave superconductor may become comparable
or even stronger than for junctions with isotropic $s$-wave
superconductors.
The main reason for this are
contributions from quasiparticle states bound to the barrier plane.

If in the vicinity of a given voltage
the dominating contribution to the conductance comes from bound states,
one may consider only a singular part (a pole-like term) of the
retarded Green function $g_S^R$ at the interface:
\begin{equation}
g_S^R\left( \bbox{ p}_{f},x=0;\varepsilon \right) =
\frac{Q_g\left( \bbox{ p}_{f}\right) }{
\varepsilon -
\varepsilon_B(\bbox{ p}_{f}) +
i\eta }
\enspace , \quad
\eta\rightarrow +0 \enspace .
\label{gprop}
\end{equation}
This yields a $\delta$-peak in the angle resolved density of states
(\ref{dos}),
\begin{equation}
\nu\left( \bbox{ p}_{f},x=0;\varepsilon \right) =
Q_g\left( \bbox{ p}_{f}\right)
\delta(\varepsilon - \varepsilon_B(\bbox{ p}_{f}))
\enspace .
\label{dospeak}
\end{equation}
Hence, the function
$\varepsilon_B(\bbox{ p}_{f})$
describes the energy dispersion of the quasiparticles
bound to the barrier.
Note that
$\varepsilon_B(\bbox{ p}_{f})$
always has an extremum for the momentum direction along
the normal of the boundary plane ($\bbox{p}_{f} \| \bbox{\hat n}$),
provided the bound state occurs for this direction
(see (\ref{epd}), Fig.\ 2 and Ref.\ \onlinecite{bs1}).
Substituting
(\ref{dospeak}) into (\ref{g})
we obtain the following anomalous contribution to the conductance:
\begin{equation}
G_s(V,T)=\frac{e^2 A}{2T}
\int\limits_{{\rm v}_{f,x}^l>0}\frac{d^{\,2}S^l}{(2\pi )^3}
\frac{{\rm v}_{f,x}^l}{|\bbox{\rm v}_{f}^l|}{D}
\left (\bbox{ p}_{f}^l\right )
\frac{Q_g\left (\bbox{ p}_{f}^l\right)}{\cosh^{2}\left (
\frac{\varepsilon_B(\bbox{p}_{f}^l)-eV}{2T}\right )}
\enspace .
\label{G}
\end{equation}
Consequently,
the conductance rapidly increases
with decreasing temperature, if the voltage is equal to
the extremum of
$\varepsilon_B(\bbox{p}_{f})$.
In case of a real order parameter bound states always appear in
pairs with energies $\pm\varepsilon_B(\bbox{p}_f)$
(see Eqs.\ (\ref{pe}), (\ref{bac})). Thus,
we get from Eq.\ (\ref{G}) the relation $G_s(V)=G_s(-V)$.

Let us again consider a two-dimensional system.
We assume that the dispersion of the bound states
can be approximated around its maximum
by
$\varepsilon_B(\bbox{s})=\varepsilon_B^{max} - bs_\|^2$
($b>0$).
Then for
$|\varepsilon_B^{max}-|eV||\ll T$
we obtain from Eq.\ (\ref{G})
the following low-temperature anomaly in the conductance:
\begin{equation}
G_s=\frac{\left(2\sqrt{2}-1\right)\zeta\left(\frac{3}{2}\right)e^2 A}{2
\sqrt{\pi bT\phantom{a} }}
\int\frac{ds_\perp}{(2\pi)^3}
\left|\frac{\partial \bbox{p}_f^l}{\partial s_{\|}}\right|
\left|\frac{\partial \bbox{p}_f^l}{\partial s_{\perp}}\right|
\frac{{\rm v}_{f,x}^l}{|\bbox{\rm v}_{f}^l|}{D}
\left (\bbox{ p}_{f}^l\right ) Q_g\left (\bbox{ p}_{f}^l\right )
\enspace.
\label{gs1}
\end{equation}

It is essential for the derivation of this result, that under the
condition
$|\varepsilon_B^{max}-|eV||\ll T$
only a region
$bs_\|^2\lesssim T$
contributes to the integral over
$s_\|$
(due to the exponential decrease of the $\cosh^{-2}$-function outside),
and the condition
$T\ll bs_{\|, max}^2$
permits to carry out the integration to infinity.
Thus, making use of only the first two terms in
the expansion of
$\varepsilon_B(\bbox{ p}_{f})$
near its maximum
may be justified at sufficiently low temperatures.
It is assumed for simplicity, that
for all values
$bs_\|^2\lesssim T$
the bound state still exists.
All other quantities are supposed to change
only slightly
within the region
$bs_{\|}^2\lesssim T$.

Provided
$T\ll |\varepsilon_B^{max}-|eV|\,|,bs_{\|,max}^2$
we obtain instead of (\ref{gs1})
the following dependence of the conductance on the voltage near the
low-temperature peak:
\begin{equation}
G_s=\frac {2e^2 A}{\sqrt{b}}
\frac{\Theta(\varepsilon_B^{max}-|eV|)}{\sqrt{\varepsilon_B^{max}-|eV|}}
\int \frac{ds_\perp}{ (2\pi)^3}
\left|\frac{\partial \bbox{p}_f^l}{\partial s_{\|}}\right|
\left|\frac{\partial \bbox{p}_f^l}{\partial s_{\perp}}\right|
\frac{{\rm v}_{f,x}^l}{|\bbox{\rm v}_{f}^l|}
{D} (\bbox{p}_{f}^l)Q_g(\bbox{p}_{f}^l)\enspace .
\label{gs2}
\end{equation}

In the case of a three-dimensional Fermi surface
we expect the function
$\varepsilon_B(\bbox{ p}_{f})$
to vary around its maximum as
$\varepsilon_B(\bbox{ p}_{f})=\varepsilon_B^{max} - b_1s_1^2 - b_2s_2^2$
$(b_1,b_2>0)$.
In contrast to the two-dimensional case
$\varepsilon_B(\bbox{ p}_{f})$
now has an extremal point instead of an extremal line.
Then Eq.\ (\ref{gs1}) is valid only
under the condition
$b_2s^2_{2,max}\ll b_1s^2_{1,max}$
(e.g., for a strongly elongated ellipsoidal Fermi surface)
for temperatures
$|\varepsilon_B^{max}-|eV||$, $b_2s^2_{2,max}\ll T\ll b_1s^2_{1,max}$.
For lower temperatures
$T\lesssim b_2s^2_{2,max}$
there are no temperature anomalies in the conductance.

Due to the dispersionless character of midgap states
($\varepsilon_B(\bbox{p}_f)=0$) their influence on the
differential conductance differs from the influence of
nonzero-energy states.
We find from (\ref{G})
\begin{equation}
G_s=\frac{e^2 A}{2T}
\int\limits_{{\rm v}_{f,x}^l>0}\frac{d^{\,2}S^l}{(2\pi )^3}
\frac{{\rm v}_{f,x}^l}{|\bbox{\rm v}_{f}^l|}{D}
\left (\bbox{ p}_{f}^l\right )
Q_g\left (\bbox{ p}_{f}^l\right )
\enspace ,
\quad |eV|\ll T
\enspace ,
\label{gmid}
\end{equation}
which is independent on the exact form and the dimensionality of
the Fermi surface. The general analytical expression for the
function $Q_g\left (\bbox{ p}_{f}^l\right )$ for midgap states via the
spatially dependent order parameter as well as the
explicit form of this function for a particular surface configuration
are given in Ref.\ \onlinecite{bs1}.
In realistic systems the width of this
zero-bias anomaly of the conductance is governed,
apart from temperature, by the width of the broadened midgap
delta-peak in the density of states.

\subsection{Numerical Results}

We have carried out numerical calculations of the conductance under
various conditions. 
The angular dependence of the transparency is taken to be
$D(\phi)=D_0\cos^2(\phi)$ where $D_0$ is the transparency for 
perpendicular incidence and is related to the normal state resistance
via Eq.\ (\ref{resist}).
The conductance $G(V)$ at zero temperature
is displayed in Fig.\ 6
for various misorientation angles $\alpha$.
Figure 6a shows the results for a smooth surface ($\rho=0$),
whereas in Figs.\ 6b, c the roughness parameter $\rho$ is chosen to
be $0.02$ and $0.10$, respectively.
For $\alpha=0$ there is no surface pair breaking and no quasiparticle
states bound to the interface.
In the case of an ideal boundary (Fig.\ 6a) the
corresponding conductance has a logarithmic divergence
(see Eq.\ (\ref{Gsing})) at $|eV|=\Delta_0$ from both sides of this
point. In contrast to this peak, one can see
asymmetrical peaks for $\alpha \ne 0$.
For finite misorientation angles $\alpha$
bound states, localized near the surface, arise. The conductance has
a square root divergence at $|eV|=\varepsilon_B^{max}$
(see Eq.\ (\ref{gs2}))
as well as a divergence at $V=0$ due to midgap states.
Surface roughness (Figs.\ 6b, c) cuts these divergences and
leads to finite peaks
which become smaller and wider with increasing roughness.
It is evident from these figures that the conductance peak
at nonzero voltage does not measure the maximum gap $\Delta_{max}$
but, to a good approximation, the (bulk) gap for perpendicular
incidence $\Delta_\infty(\bbox{p}_f^n)$.
We want to stress that the position of the peaks changes only
slightly with increasing roughness.
Although a direct comparison of our data with the results of
Ref.\ \onlinecite{nag96b} is not possible (they
consider the surface density of states and not the conductance)
the influence of roughness is similar in both models.

The low-temperature behavior of the conductance peaks
is presented in Figs.\ 7a, b
for smooth interfaces ($\rho=0$)
with misorientation angles $\alpha=20^\circ$ and $30^\circ$.
The temperature dependence
of the zero-bias anomaly $G(0)$ is described in Fig.\ 7a,
while in Fig.\ 7b the values of the peak at nonzero voltage
$G(V_{max})$ are shown. We see, for example,
that for a lattice to surface orientation $\alpha=20^\circ$
the zero-bias conductance $G(0)$ is proportional with a good accuracy to
$1/T$ up to $T=0.2\, T_c$, whereas the temperature dependence
of the quantity $G(V_{max})$ is described by
the $\left(1/\sqrt{T}\right)$--term only for a very small
temperature range ($T\lesssim 0.01\, T_c$).

The influence of finite temperatures on the conductance is plotted
in Fig.\ 8 (Fig.\ 9) for a lattice to surface orientation
$\alpha=20^\circ$ ($\alpha=45^\circ$).
We present results for different degrees of interface roughness:
(a) smooth interface ($\rho=0$),
(b) slightly rough interface ($\rho=0.02$) and
(c) higher degree of roughness ($\rho=0.10$).
The peaks are well pronounced only for $T \lesssim 0.2\, T_c$.
Even at $T=0.05 \, T_c$ (this corresponds to the temperature
of liquid Helium if $T_c=93$ K is assumed) the reduction of
the height of the conductance peaks is evident.
At least at low temperatures, the effect of thermal smearing is
more pronounced for the nonzero-bias peaks (Fig.\ 9a).
Thus, we would like to encourage experimentalists to do
tunneling measurements even below $4.2$ K.

Figure 10 compares the results for the conductance at $T=0$
using a constant order parameter (cf.\ Ref.\ \onlinecite{tan3})
with calculations including a self-consistent order parameter.
In both cases ($\alpha=20^\circ$ and $\alpha=45^\circ$)
we find dramatic differences at finite voltages because of the
appearance of nonzero-energy bound states due to the depletion of
the order parameter near the interface.
The presence of the zero-bias conductance peak is unaffected by the
self-consistency of the order parameter, although disregarding
surface pair-breaking results in overestimating of the
weight of the peak up to 50\% depending on the crystal orientation.

\subsection{Summary}

In summary, we have studied in detail the appearance of
surface bound states
and their relevance for the tunneling spectrum of NIS junctions.
We developed, for the first time, an analytical approach to the
surface problem which allows an understanding of the presence of
zero-energy as well as nonzero-energy bound states.
Numerical calculations have been done for $d$-wave superconductors
assuming a cylindrical Fermi surface.

It is well known that zero-energy bound states are
a robust phenomenon\cite{ati75}
and will always occur whenever the real order parameter changes sign
along a quasiparticle trajectory.
So far, it was not clear under which condition
nonzero-energy bound states arise. They were only observed
numerically.\cite{mat,buch95}
The result of our analytical consideration is that at least some of the
nonzero-energy bound states are also of quite general (although not
topological) origin and appear, for
any (even tiny) suppression of the order parameter, within
a narrow region around the surface normal (see Fig.\ 2a).
The reason is that the order parameter
amplitudes for the incoming and the outgoing trajectories
far away from the surface are approximately the same
in this region,
i.e., $\Delta_\infty(\bbox{p}_f) \approx
\Delta_\infty(\underline{\bbox{p}}_f)$.
As a consequence, nonzero-energy states around the surface normal
occur in the case of $d$-wave superconductors
for all lattice to surface orientations
with the exception of (100) surfaces (no suppression of the order
parameter) and of (110) surfaces ($\Delta(\bbox{p}_f^n)=0)$.
In addition, we found numerically nonzero-energy bound states
for misorientation angles between $37^\circ$ and $45^\circ$
as shown in Fig.\ 2b. These states are, however, beyond the validity
of our analytical approach.

Furthermore, we have studied the influence of bound states on
the tunneling conductance, in particular at low temperatures
(Figs.\ 6a).
While zero-energy bound states always result in a $1/T$ divergence
of the conductance for $T \rightarrow 0$,
the influence of nonzero-energy bound states
depend on the geometry of the Fermi surface.
For quasi two-dimensional Fermi surfaces the conductance
diverges as $1/\sqrt{T}$ at voltages determined
by the extrema of the bound state energies.
Within a few percent, these peaks coincide for $\alpha\lesssim 37^\circ$
with the bulk order parameter for perpendicular incidence,
i.e., $|eV_{max}|\approx|\Delta_\infty(\bbox{p}_f^n)|$.
Only for higher misorientation angles ($\alpha \gtrsim 37^\circ$)
the conductance peaks correspond approximately to the bulk gap
$\Delta_{max}=\mathop{\rm max}_{{\bf p}_f}|\Delta_\infty(\bbox{p}_f)|$.
By contrast, low-temperature peaks in the conductance at
nonzero-bias voltage are absent for three-dimensional systems.
Our finite temperature calculations (Figs.\ 8a, 9a) demonstrate
that the peaks can only be observed for temperatures below $0.2\, T_c$.

Finally, we have investigated the influence of surface roughness
on the conductance (Figs.\ 6b, c, 8b, c, 9b, c).
The peaks are broadened, but their positions
remain essentially untouched. The broadening of the peaks with
increasing surface roughness is more crucial for the zero-bias peaks.

We conclude that a systematic experimental study of low-temperature
peaks in the conductance of NIS junctions for a series of lattice to
surface orientations would give valuable information on
the anisotropy of the superconducting order parameter.
We stress once more that nonzero-bias peaks give
information on the gap amplitude in the bulk, even though the order
parameter is suppressed at the surface.

\section*{Acknowledgments}
We wish to thank D.~Rainer for many stimulating discussions
and a critical reading of the manuscript.
This work is supported in part by grant No.~96-02-16249
of the Russian Foundation for Basic Research. A.A.S.~acknowledges
financial support by the Forschungszentrum J\"ulich.

\newpage


\newpage


\begin{figure}[t]
\caption{
Schematic geometry of our interface model: a smooth interface
covered by a dirty layer on each side. Also shown are the
four Fermi momenta which are involved in the scattering process
at the smooth interface ($x=0$).
}
\end{figure}
\begin{figure}[t]
\caption{
Angular dependence of the bound state energies $\varepsilon_B(\phi)$
near $\phi=0^\circ$ (a) and near $\phi=45^\circ$ (b)
for various misorientation angles $\alpha$.
}
\end{figure}

\begin{figure}[t]
\caption{
Bound state energy for perpendicular incidence $\varepsilon_B(\phi=0)$
as a function of the misorientation angle $\alpha$.
For comparison, we included the position of the band edge,
$\Delta_\infty(\phi=0)/\Delta_{max}=\cos(2\alpha)$.
}
\end{figure}

\begin{figure}[t]
\caption{
Angle-resolved density of states at a smooth surface ($\rho=0$) for
various trajectory angles $\phi$.
The lattice to surface orientation is taken to be
$\alpha=0^\circ$ (a), $20^\circ$ (b) and $45^\circ$ (c).
}
\end{figure}

\begin{figure}[t]
\caption{
Angle-resolved density of states at rough surfaces
($\alpha=20^\circ$) for the same trajectory angles $\phi$
as in Fig.\ 4b. The roughness parameter $\rho$ is
$0.02$ (a), $0.10$ (b) and $0.50$ (c).
}
\end{figure}
\begin{figure}[t]
\caption{
Zero-temperature conductance as a function of the bias voltage
for various misorientation angles $\alpha$.
The surface roughness is $\rho=0$ (a), $0.02$ (b) and $0.10$ (c).
The inset of (a) demonstrates the presence of
three conductance peaks for $\alpha=40^\circ$.
}
\end{figure}

\begin{figure}[t]
\caption{
Temperature dependence of the zero-bias anomaly $G(V\!=\!0,T)$ (a)
and the nonzero-bias anomaly $G(V\!=\!V_{max},T)$ (b)
for two lattice to surface orientation angles
$\alpha=20^\circ$ and $30^\circ$.
}
\end{figure}
\begin{figure}[t]
\caption{
Conductance $G(V)$ for various temperatures $T$.
The lattice to surface orientation is taken to be $\alpha=20^\circ$.
The figures display the effect of increasing roughness:
$\rho=0$ (a), $0.02$ (b) and $0.10$ (c).
}
\end{figure}
\begin{figure}[t]
\caption{
The same as in Fig.\ 8 but for $\alpha=45^\circ$.
}
\end{figure}

\begin{figure}[t]
\caption{
Influence of the self-consistency of the order parameter on the
zero-temperature conductance of a smooth interface ($\rho=0$).
We show the data for two misorientation angles
$\alpha=20^\circ$ and $45^\circ$.
}
\end{figure}




\begin{references}
\bibitem{not} We follow the notation of Ref.\ \onlinecite{buch95}.
\end{references}
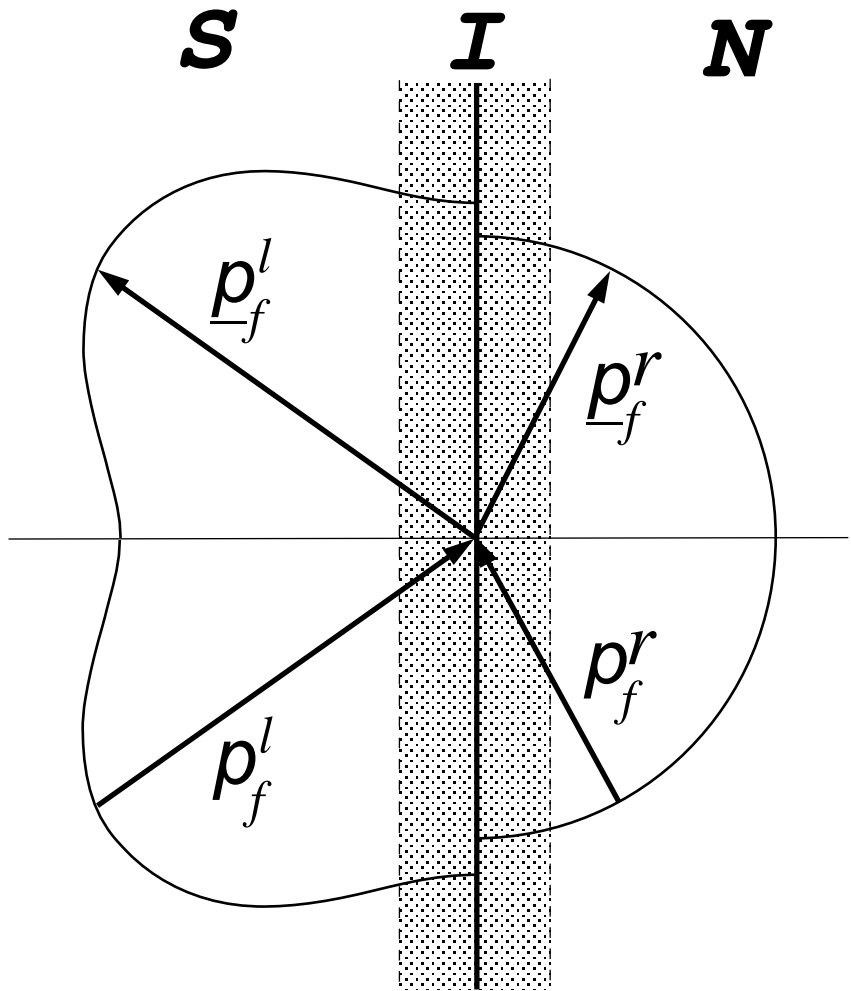
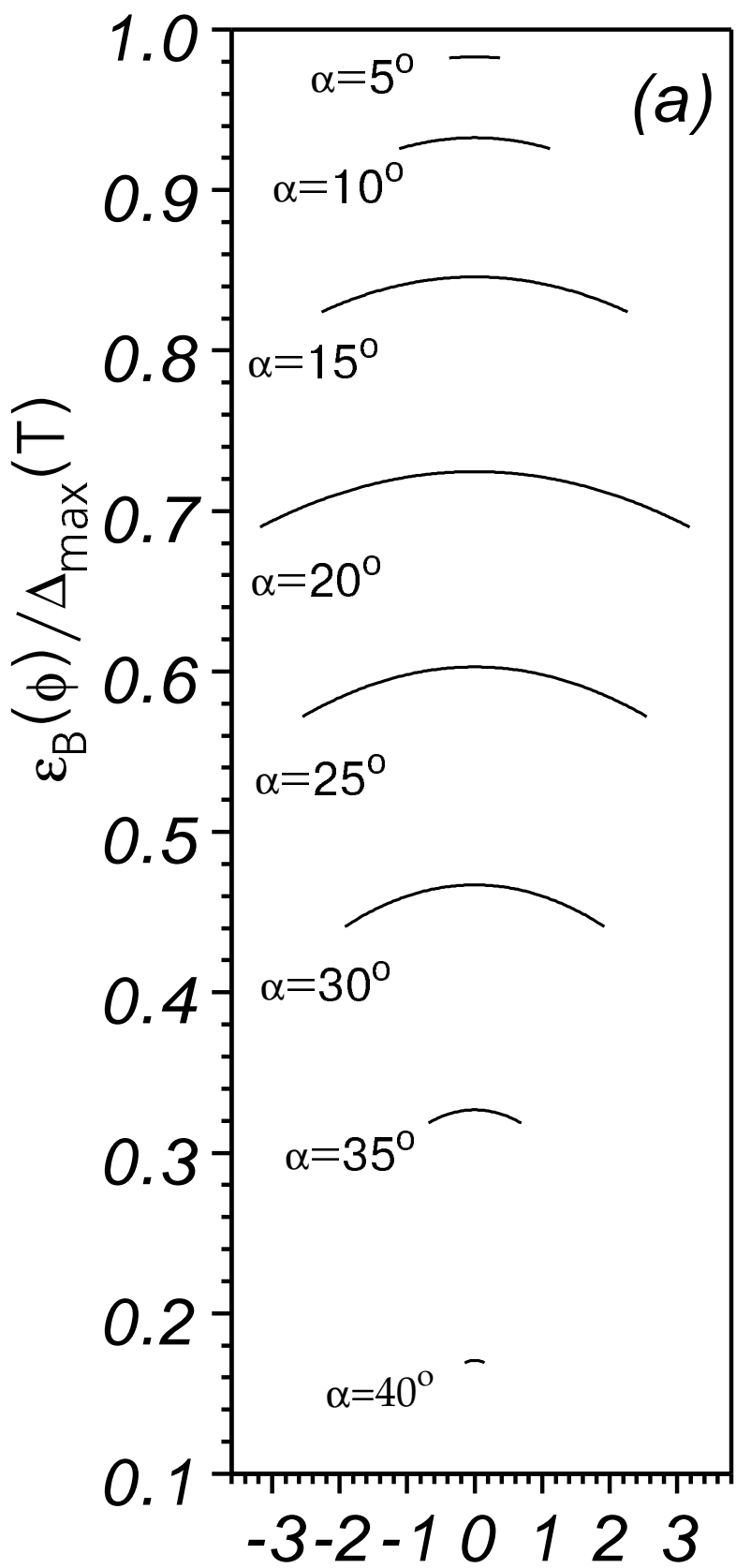
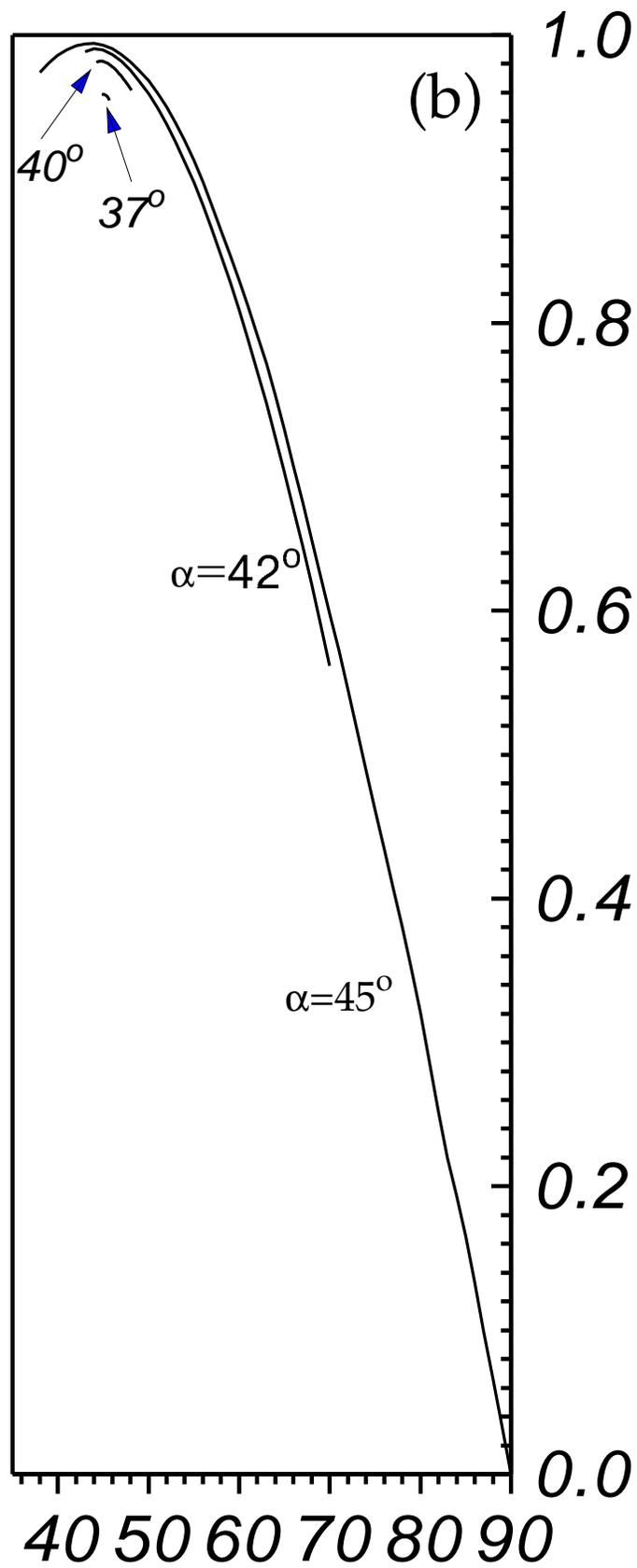
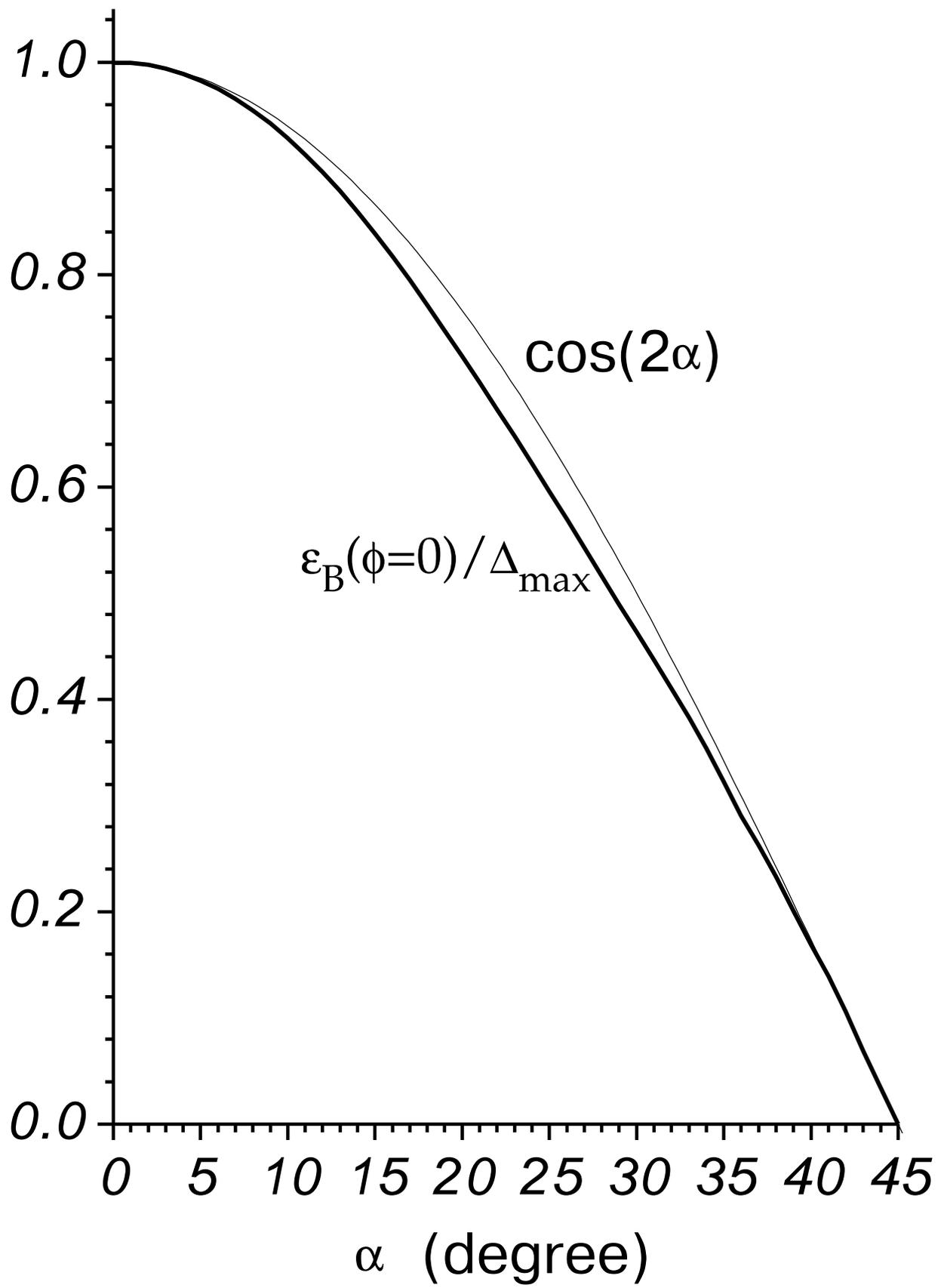
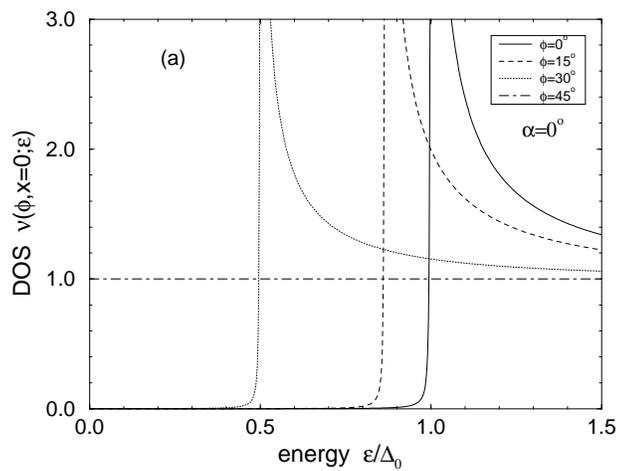
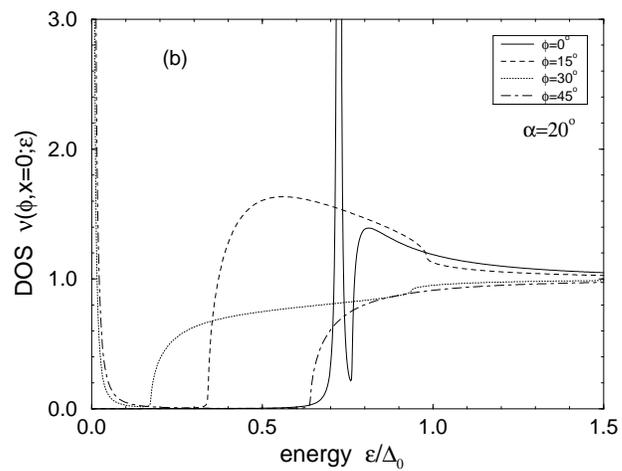
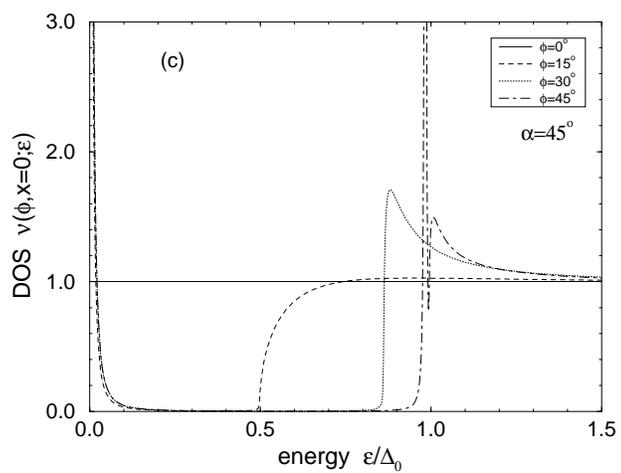
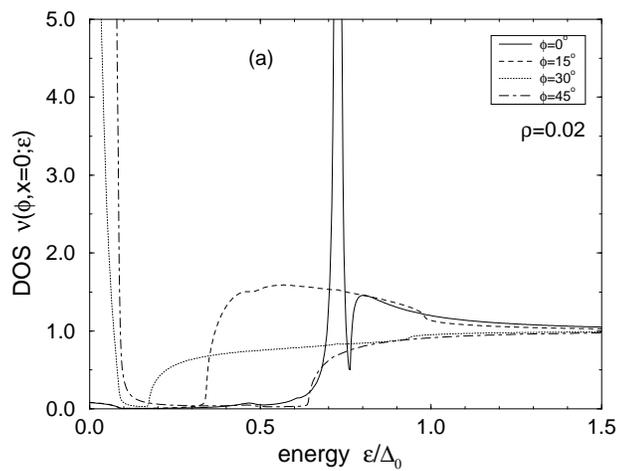
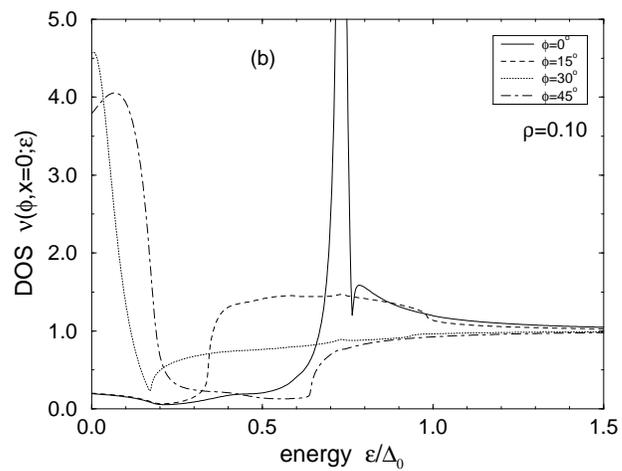
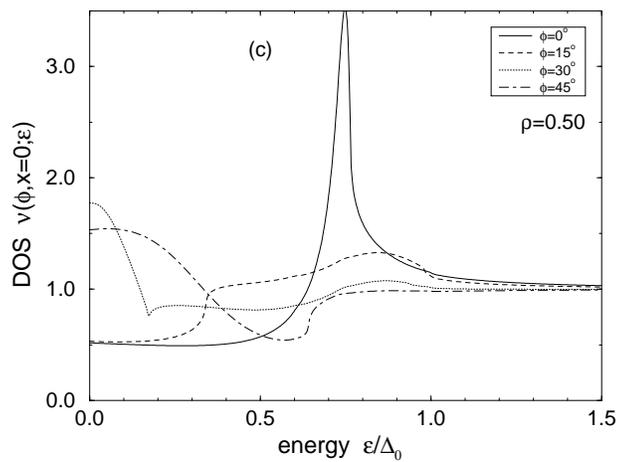
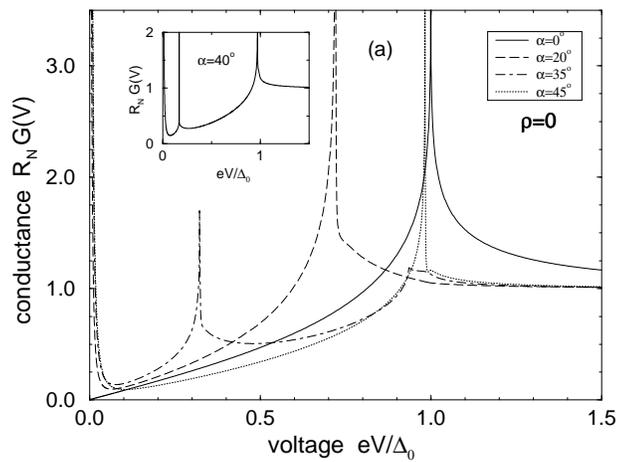
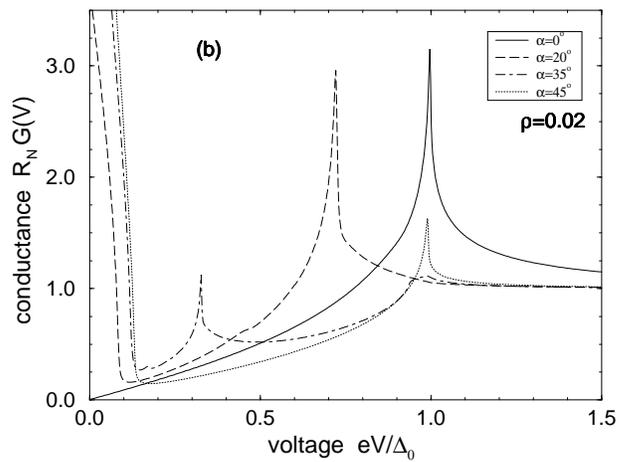
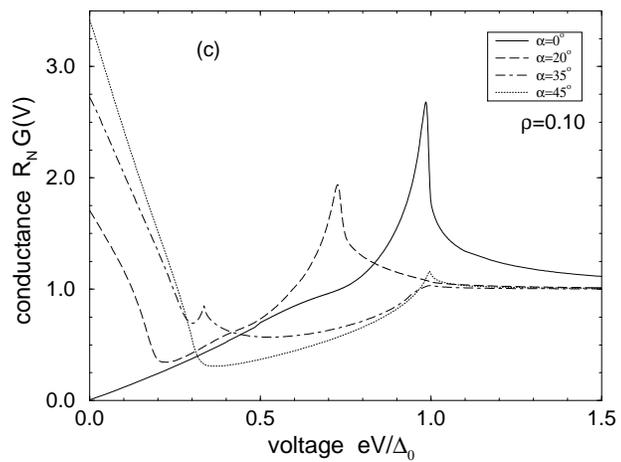
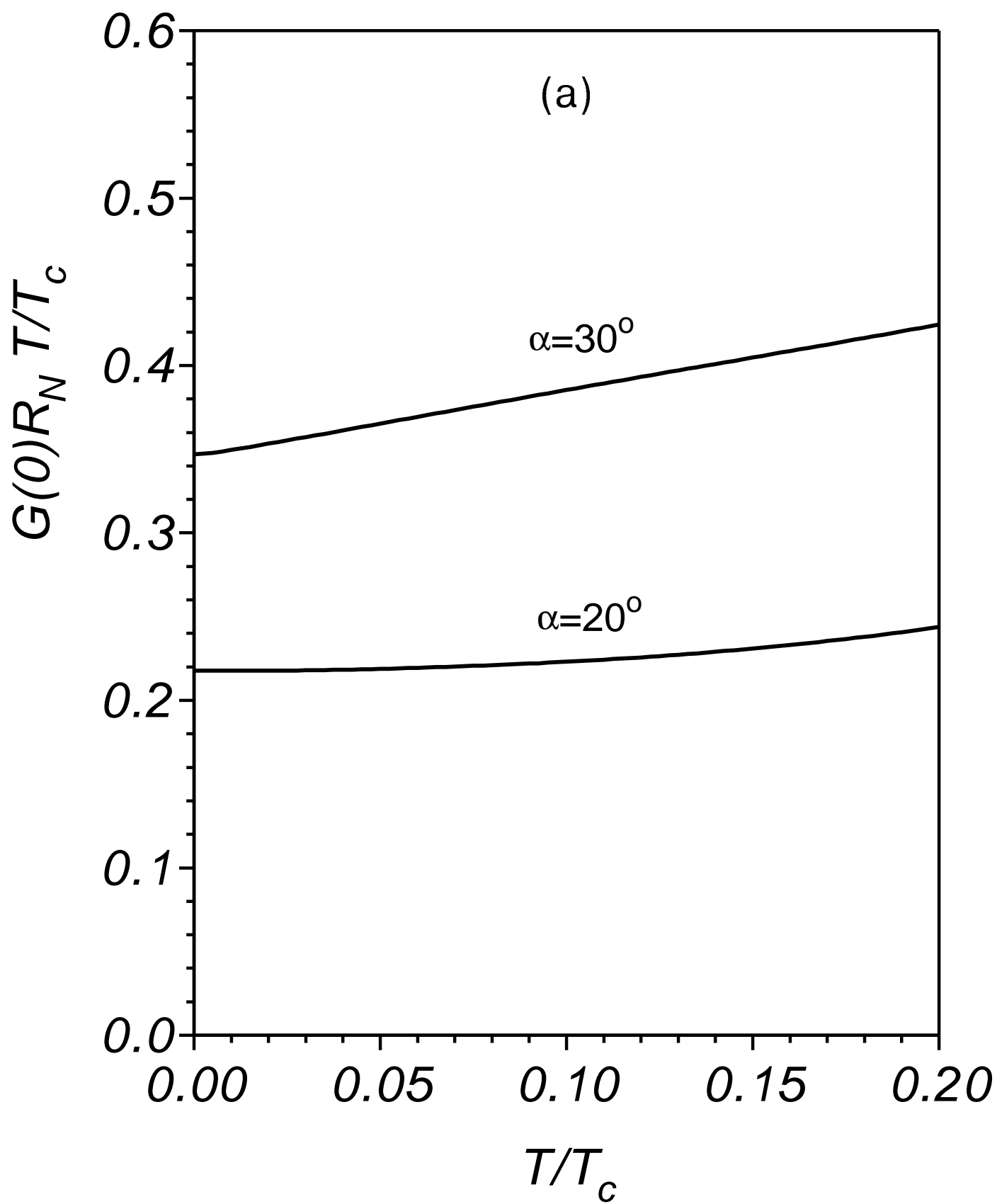
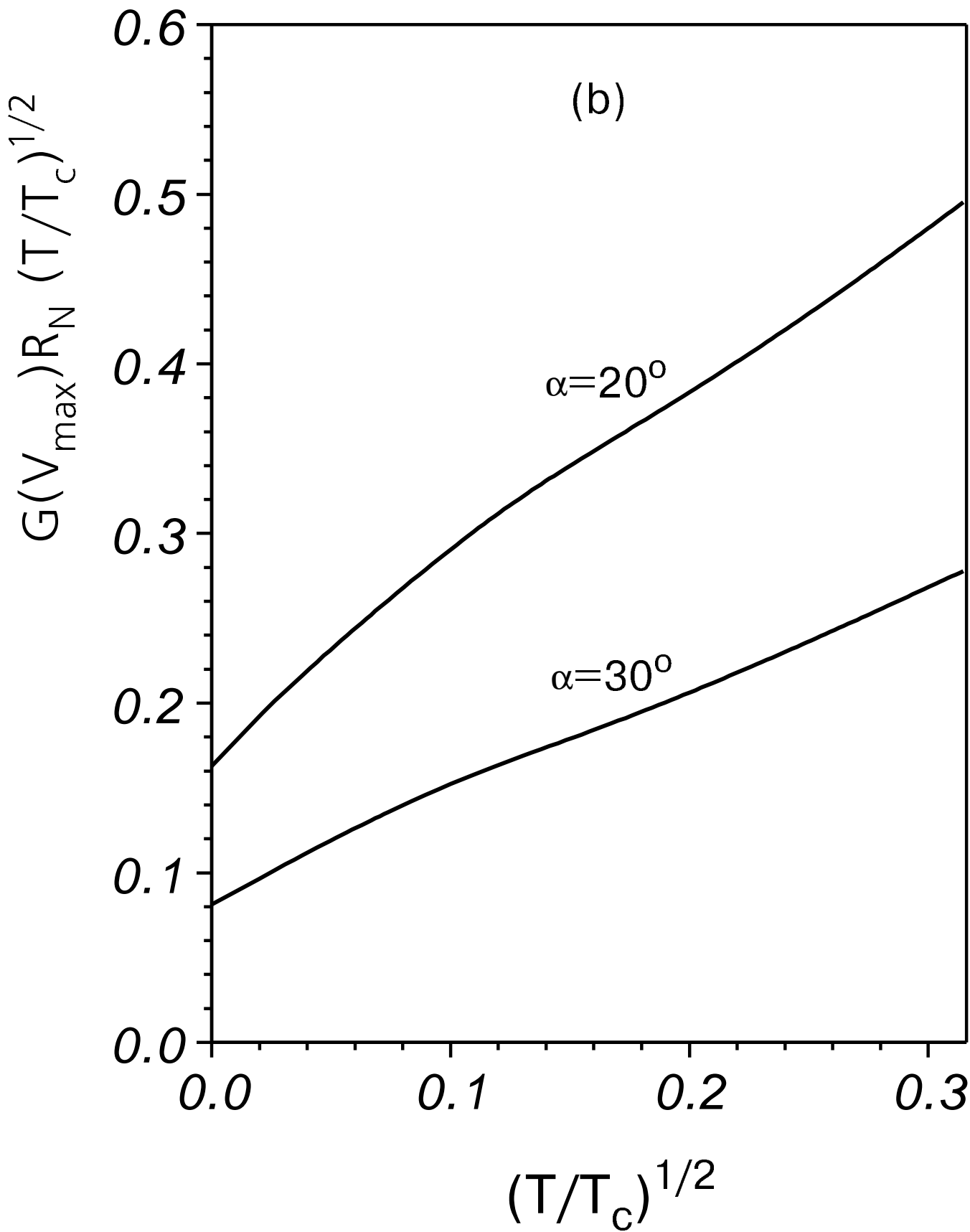
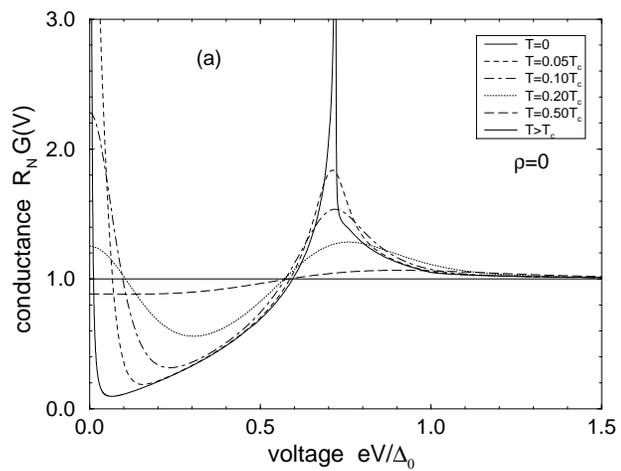
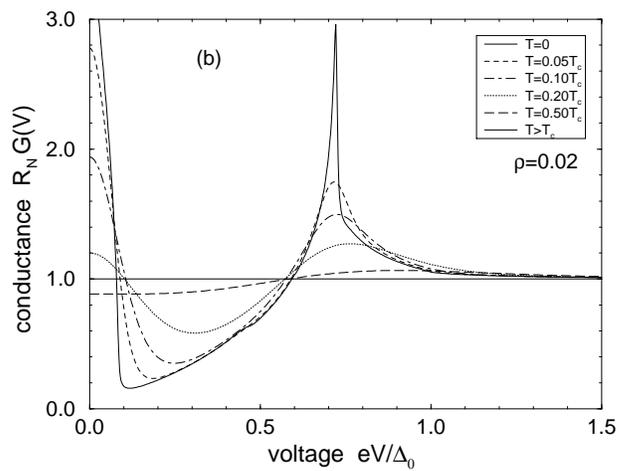
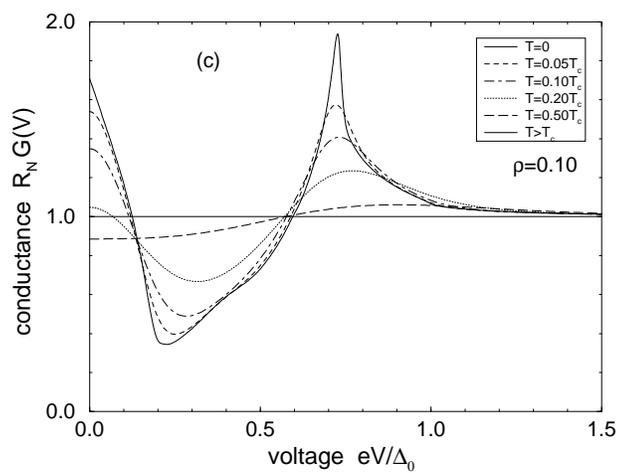
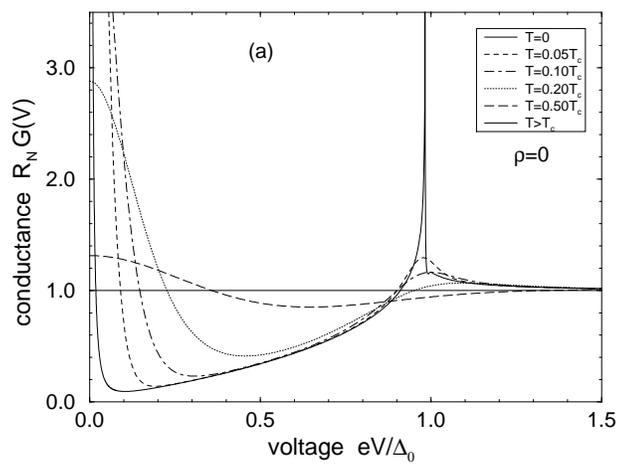
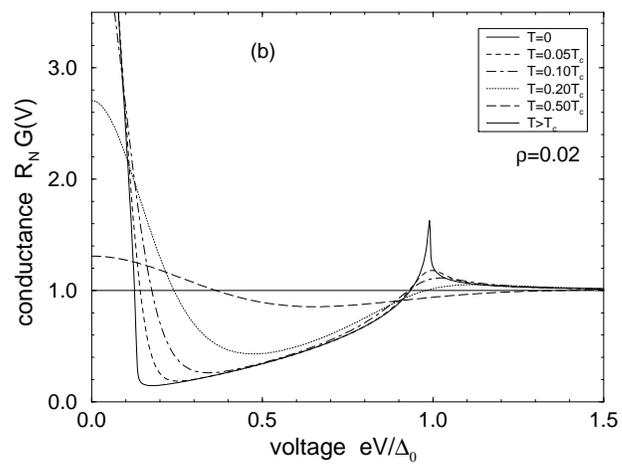
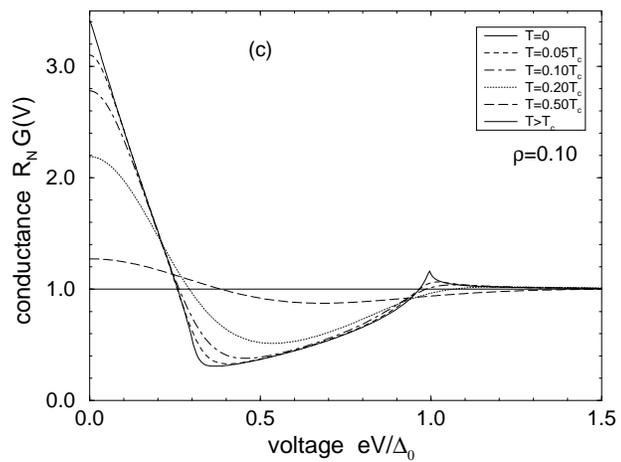
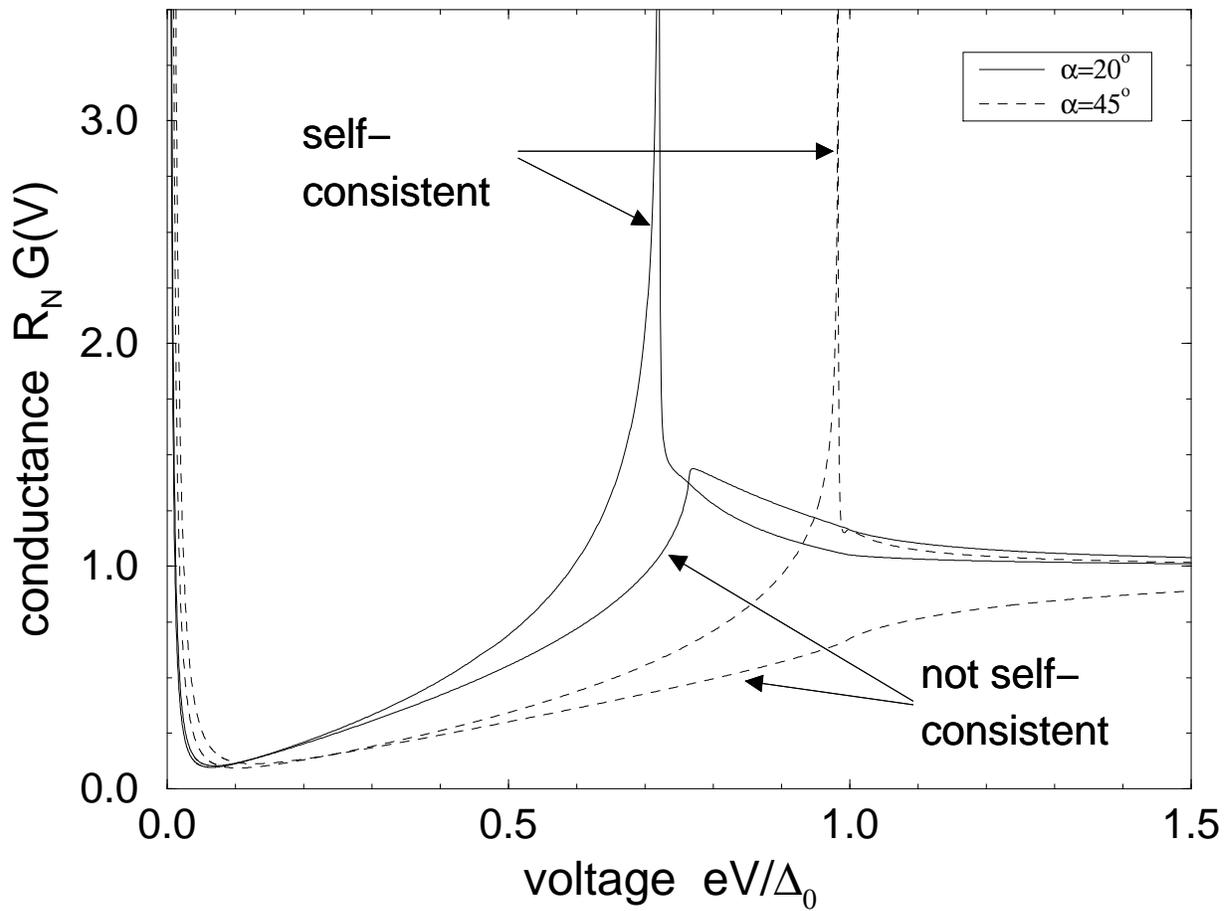
\end{document}